\documentclass[spanish,english]{article}
\usepackage[T1]{fontenc}
\usepackage[latin9]{inputenc}
\usepackage{geometry}
\usepackage{float}
\usepackage{mathrsfs}
\usepackage{amsmath}
\usepackage{amssymb}
\usepackage{graphicx}
\usepackage[all]{xy}

\makeatletter
\newenvironment{lyxlist}[1]
	{\begin{list}{}
		{\settowidth{\labelwidth}{#1}
		 \setlength{\leftmargin}{\labelwidth}
		 \addtolength{\leftmargin}{\labelsep}
		 }}
	{\end{list}}

\makeatother

\usepackage{babel}
\addto\shorthandsspanish{\spanishdeactivate{~<>}}

\begin{document}
\title{The Spacetime Picture in Quantum Gravity II}
\author{{\normalsize{}Alejandro Ascárate}\thanks{FaMAF-UNC (Córdoba, Argentina). Email: aleazk@gmail.com}}
\maketitle
\begin{abstract}
{\normalsize{}As a continuation to \cite{key-12}, we introduce a
new formalism for (part of) QG, which we call TQG, since it's based
on the NC Tori. This allows us to obtain numerous insights about the
nature of time, like its discretization, its regular pace at the macroscopic
scale, a solution to the Problem of Time, and a connection with the
Measurement Problem and wave function collapse.}{\normalsize\par}
\end{abstract}
$\;$

\subsection*{{\normalsize{}Introduction}}

$\;$

In the phase space of Gravity, there's no background spacetime (both
in terms of manifold and metric), and this leads to a Hamiltonian
$\mathcal{H}_{EH}=\sum_{i}N_{(i)}\mathbf{C}^{(i)}$ given by the constraints
$\mathbf{C}^{(i)}$, which, in turn, leads to gauge invariant phase
space properties (``Dirac properties'') $F$ (i.e., the ones that
commute with the constraints, $\left\{ F,\mathbf{C}^{(i)}\right\} =0$)
that lack an adequate time evolution (since $\frac{\partial F}{\partial t}=\left\{ F,\mathcal{H}_{EH}\right\} =$$\sum_{i}N_{(i)}\left\{ F,\mathbf{C}^{(i)}\right\} =0$.)
We claim that Dirac properties are not the relevant thing to look
for. For example, one can show that the total volume of a solution
$g$ is not a Dirac property, even when it's a diffeomorphism invariant
functional \cite{key-2}. But the problem is not in the volume, it's
in the notion of Dirac property. Indeed, the reason for the previous
seemingly paradoxical issue is directly related to the fact that the
constraints in GR do not implement the full spacetime diffeomorphism
group on phase space, and this is a consequence of the fact that the
slicing of spacetime with spacelike Cauchy surfaces (which is needed
to build the phase space) is dependent on the dynamical metric \cite{key-2}.
Thus, Dirac properties are phase space properties, and very tied to
its structure. The issue about diffeomorphism invariant properties
in a solution, but which are not Dirac properties, is, again, an artifact
of the phase space picture, like the ``no time'' problem. To avoid
all of these problems, we propose to completely dispense from the
phase space picture when dealing with geometrical properties of spacetime,
and instead switch to a spacetime picture. The only reason why we
consider the Hamiltonian formulation (and, thus, the phase space picture
and Dirac properties) is because it's needed for the process of canonical
quantization. But, once we canonically quantize a kinematical algebra,
we can build the relational spacetime algebra and start to work in
the spacetime picture and thus forget about the phase space; only
if we are trapped in the phase space picture we would need to consider
Dirac properties. When we work in basic GR in the spacetime picture
and solve the field equations in simple cases, we don't even think
about Dirac properties, we just consider the diffeomorphism invariant
properties of the solution in question, like the spacetime volume
(and, of course, also the time evolution with respect to this solution;
it doesn't make sense to look for ``time evolutions in the phase
space of GR'' since there's no spacetime in it, it's only when we
pick a solution, a metric, that we can build the relational spacetime,
as was argued here: time evolution is the specific change with respect
to a \emph{duration}, the variable with respect to which one measures
the change \emph{must have that specific physical interpretation};
thus, since we need anyway to pick a solution to build the relational
spacetime and the time evolution, the need to build Dirac phase space
properties in order to consider their ``phase space time evolution''
completely dissolves, since by definition we must abandon the phase
space picture if we want to build a true time evolution; the imperative
of Dirac properties in QG comes from the assumption that we need,
and can, build a true time evolution in the phase space of GR, for
which, naturally, one would need to consider the phase space version
of diffeomorphism invariance, namely, Dirac properties, an a notion
of relational time not related to duration, a property of the gravitational
field, but to other fields, if that is even possible in the first
place.)

$\;$

Thus, the problem here will be assumed to be that of trying to find
an analogue of duration with respect to a given metric (i.e., in a
spacetime picture) but now in the quantum (gravity) realm.

$\;$

\subsection*{{\normalsize{}Background}}

$\;$

We review and summarize below some of the basic notions from \cite{key-12}
that will be relevant in the present paper, which is a continuation.
We follow here the numeration for the definitions and propositions
from that reference, where a more detailed discussion can be found.

$\;$

$\mathbf{Definition\;1.1}$: The (kinematical) phase space of GR is
defined as

\[
X=\left\{ \left[h_{ab},\pi^{ab}\right]\diagup h_{ab},\pi^{ab}\in C^{\infty}(\varSigma)\,(\mathrm{as\,fields}\,\mathrm{on}\,\varSigma)\right\} ,
\]
where $h_{ab}$ and $\pi^{ab}$ are, respectively, a smooth riemannian
metric on a spacelike Cauchy hypersurface $\Sigma$ in a compact and
boundaryless spacetime $M$, foliated by $\Sigma$ as usual, and the
conjugate momentum tensor density$.\,\blacksquare$

$\;$

$\mathbf{Definition\;1.2}$: The subset $\mathcal{F}\subset C(X)$
consists of the phase space functionals of the form

\[
F_{f}\left(\left[h_{ab},\pi^{ab}\right]\right)\doteq\int_{\varSigma}f\boldsymbol{\epsilon}(h_{ab}),\,\forall\left[h_{ab},\pi^{ab}\right]\in X,
\]
where $f\in C^{\infty}(\varSigma)$ and $\boldsymbol{\epsilon}(h_{ab})$
is the volume element of $h_{ab}$, i.e. $\boldsymbol{\epsilon}(h_{ab})=\sqrt{h}\,\mathrm{d}^{3}x$
($h=\mathrm{det}\,(h_{ab})$)$.\,\blacksquare$

$\;$

$\mathbf{Proposition\;1.1}$: The assignment $f\,\longmapsto F_{f}$
is injective.$\,\square$

$\;$

$\mathbf{Definition\;1.3}$: We now define a mapping $R:\mathcal{F}\times X\,\longrightarrow C^{\infty}(\varSigma)\times Obj(\mathsf{Hil})$
(where $Obj(\mathsf{Hil})$ is the collection of objects in the category
$\mathsf{Hil}$ of Hilbert spaces) by 
\[
(F_{f},\left[h_{ab},\pi^{ab}\right])\,\longmapsto R(F_{f},\left[h_{ab},\pi^{ab}\right])\doteq(f,L^{2}(\mathcal{S},\boldsymbol{\epsilon}(h_{ab})),
\]
(where $\mathcal{S}$ is the $C^{\infty}(\varSigma)-$module of smooth
spinor fields in $\varSigma$, which from now on we assume allows
a spin structure)$.\,\blacksquare$

$\;$

$\mathbf{Definition\;1.4}$: For \emph{fixed} $\left[h_{ab},\pi^{ab}\right]\in X$
and \emph{variable} $F_{f}\in\mathcal{F}$, we denote the first component
of $R(F_{f},\left[h_{ab},\pi^{ab}\right])$ as $R_{h}(F_{f}).\,\blacksquare$

$\;$

$\mathbf{Definition\;1.5}$: We make $\mathcal{F}$ into an (unital)
algebra $(\mathcal{F},$$\cdot_{sp.})$, where the product $\cdot_{sp.}$
in $\mathcal{F}$ is defined as

\[
\left(F_{f_{1}}\cdot_{sp}F_{f_{2}}\right)\left(\left[h_{ab},\pi^{ab}\right]\right)\doteq F_{f_{1}f_{2}}\left(\left[h_{ab},\pi^{ab}\right]\right),\,\forall\left[h_{ab},\pi^{ab}\right]\in X
\]

\[
=\int_{\varSigma}f_{1}f_{2}\boldsymbol{\epsilon}(h_{ab}).\,\blacksquare
\]

$\;$

$\mathbf{Corollary\;1.1}$: The map $R_{h}$ is a \emph{bijection}
beetwen $\mathcal{F}$ and $C^{\infty}(\varSigma)$ (by Proposition
1.1), and is a faithfull algebra representation of $\mathcal{F}$
into\footnote{$\mathcal{B}(\mathcal{H})$ is the algebra of all the \emph{bounded}
operators that act on the Hilbert space $\mathcal{H}$, with the operator
composition $\circ$ as algebraic product.} $\mathcal{B}\left(L^{2}(\mathcal{S},\boldsymbol{\epsilon}(h_{ab}))\right)$
by multiplication operators, i.e. $R_{h}(F_{f})(\psi)(x)\doteq f(x)\psi(x),\,\forall\psi\in L^{2}(\mathcal{S},\boldsymbol{\epsilon}(h_{ab}))$$.\,\square$

$\;$

We call $R_{h}$ the ``\emph{relational representation}'', since
\emph{it gives a way of obtaining the algebra of physical $3-$space
purely from Field properties} (represented here by the phase space
functionals). \emph{Switching to a relational and algebraic frame
of mind, we can take this representation as the actual way of defining
how to build space from the phase space algebra of the gravitational
field.}

$\;$

$\mathbf{Proposition\;1.2}$: If one starts with a subset $\mathcal{A}_{\varSigma}$
(with a commutative product $\cdot_{sp}$) of the phase space algebra
$\mathcal{A}_{X}$ (i.e. $\mathcal{A}_{\varSigma}\subset\mathcal{A}_{X}$,
but only as a set and\emph{ not} as a subalgebra\footnote{This convention will be maintained whenever the symbol $\subset$
is used, unless some other sense is explicitly stated.}), which correspond, respectively, to $\mathcal{F}$ and $C(X)$ but\emph{
seen as abstract algebras}, then there exist a \emph{purely algebraic}
(i.e. which \emph{doesn't }make use of the commutative manifold structure
of space for its definition) faithful representation $\left(\widetilde{R}_{h}\left[\mathcal{A}_{\varSigma}\right],\mathcal{H}_{\widetilde{R}_{h}}\right)$
of $\mathcal{A}_{\varSigma}$ and isomorphisms $\pi_{X}:\mathcal{A}_{\varSigma}\,\longrightarrow\mathcal{F}$
and $\pi_{\varSigma}:\widetilde{R}_{h}\left[\mathcal{A}_{\varSigma}\right]\,\longrightarrow C^{\infty}(\varSigma)$,
such that the following diagram commutes:

\[
\xymatrix{\mathcal{A}_{\varSigma}\ar[d]^{\widetilde{R}_{h}}\ar[r]_{\pi_{X}} & \mathcal{F}\ar[d]_{R_{h}}\\
\widetilde{R}_{h}\left[\mathcal{A}_{\varSigma}\right]\ar[r]^{\pi_{\varSigma}} & C^{\infty}\left(\varSigma\right)
}
\]
i.e. such that

\[
R_{h}=\pi_{\varSigma}\circ\widetilde{R}_{h}\circ\pi_{X}^{-1}.\,\square
\]

$\;$

Note that,\emph{ in the general, non-commutative case, only the left
hand side of the previous diagram survives}. In this way, \emph{all
the geometrical information of the space is now contained in the spectral
triple}.

$\text{\;}$

$\mathbf{Definition\;2.1}$: Consider the phase space of GR, $(X,C(X),\left\{ \cdot,\cdot\right\} _{GR})$,
and a background independent Poisson sub-algebra $\mathcal{A}_{GR}$
of phase space functionals there. The basic $^{*}-$algebra, $\widehat{\mathcal{A}}_{QG}$,
in QG will be the algebra $S_{\mathcal{A}_{GR}}$ freely generated
(with field $\mathbb{C}$) by the classical algebra elements\footnote{Which is just the complex vector space generated by the basis $\left\{ e_{S}\right\} $,
where $S$ runs over all the possible ordered, finite sequences of
elements from $\mathcal{A}_{GR}$ (e.g. $S=$$(a_{1},a_{2},...,a_{k}),\,k>0$)
and the algebra product is given at the basis level by $e_{S}\star e_{T}\doteq e_{(S,T)}$.} and with the relations imposed by the ideals $\mathcal{I}_{*LP}$
(where $*$ is the involution, $L$ is from linear and $P$ from $\mathrm{i}$
multiplied by the Poisson brackets. This process simply imposes the
familiar Dirac commutation relations of canonical quantization to
$\widehat{\mathcal{A}}_{QG}$, that is: $\left[\widehat{a},\widehat{a}'\right]_{\widehat{\mathcal{A}}_{QG}}\doteq$$\widehat{a}\star_{\widehat{\mathcal{A}}_{QG}}\widehat{a}'-\widehat{a}'\star_{\widehat{\mathcal{A}}_{QG}}\widehat{a}=$$\mathrm{i}\widehat{\left\{ a,a'\right\} }_{GR}$,
where $\star_{\widehat{\mathcal{A}}_{QG}}$ is the quantum probability
algebraic product of $\widehat{\mathcal{A}}_{QG}$). That is, $\widehat{\mathcal{A}}_{QG}$
is the quotient:

\[
\widehat{\mathcal{A}}_{QG}\doteq\frac{S_{\mathcal{A}_{GR}}}{\mathcal{I}_{*LP}}.\,\blacksquare
\]

$\;$

The next task is to identify a \emph{subalgebra} $\widehat{\mathcal{A}}_{SP.}\subset\widehat{\mathcal{A}}_{QG}$\emph{
that plays a role equivalent to the one of $\mathcal{A}_{\varSigma}$
in the classical case of Proposition 1.2.}

$\;$

$\mathbf{Definition\;2.2}$: Consider the subset $\{f_{S}\}$ of all
elements $f_{S}$ in $\mathcal{F}$ with $S\doteq supp\,f_{S}\subseteq\varSigma$,
then we can see $\mathcal{F}$ as the result of the application of
an \emph{indexation relation} $\widetilde{I}_{\mathcal{F}}:I\,\longrightarrow\mathcal{F},\,S\,\longmapsto\widetilde{I}_{\mathcal{F}}(S)\doteq\{f_{S}\}$,
where $I$ is the collection of all the supports $S$. We now \emph{replace}
$I$ by a subset $I_{Dis}\subset I$ which can be \emph{at most countably
infinite}, and compute the Poisson brackets in $\mathcal{F}_{Dis}\doteq\widetilde{I}_{\mathcal{F}}[I_{Dis}]$:
the subset $I_{Dis}$ should be selected in a way that makes $(\mathcal{F}_{Dis},\left\{ \cdot,\cdot\right\} _{GR})$
a Poisson subalgebra and, in particular, one whose structure constants
\emph{don't }depend on the differentiable manifold details of $\Sigma$
nor on the details of the functions $f_{S}$ varying over it, \emph{but
only on $I_{Dis}$, as an indexing set.} With this set up, we choose
$\mathcal{A}_{GR}$ such that $\mathcal{F}_{Dis}\subset\mathcal{A}_{GR}$
(so that $\widehat{\mathcal{F}}_{Dis}\subset\widehat{\mathcal{A}}_{QG}$)
and define:

\[
\widehat{\mathcal{A}}_{SP.}\doteq\widehat{\mathcal{F}}_{Dis}.\,\blacksquare
\]

$\;$

$\mathbf{Definition\;2.3}$: Assuming one has the family of possible
representations $\left(R_{QG}\left[\widehat{\mathcal{A}}_{QG}\right],\mathcal{H}_{R_{QG}}\right)$
of $\widehat{\mathcal{A}}_{QG}$, we define as

\[
\widehat{\mathcal{A}}_{sp.}\doteq R_{QG}\left[\widehat{\mathcal{A}}_{SP.}\right]
\]
the \emph{non-commutative} algebra of quantum physical $3-$space
that ``relationally arises'' from the quantum gravitational field
(this algebra will be the quantum analogue of $\widetilde{R}_{h}\left[\mathcal{A}_{\varSigma}\right]$,
and $R_{QG}$ that of $\widetilde{R}_{h}$, in the left hand side
of the diagram of Proposition 1.2)$.\,\blacksquare$

$\;$

$\mathbf{Definition\;2.6}$:\foreignlanguage{spanish}{ The analogue
of a given classical metric $h_{ab}$ will be given by a $3-$(spectral)
dimensional \emph{real first order spectral triple} 
\[
\mathcal{T}_{D_{QG}}^{R_{QG}}\doteq\left(R_{QG}\left[\widehat{\mathcal{A}}_{SP.}\right],\mathcal{H}_{R_{QG}},D_{QG}\right).\,\blacksquare
\]
}

\selectlanguage{spanish}%
$\;$

\selectlanguage{english}%
$\mathbf{Remark\;2.2}$:\foreignlanguage{spanish}{ In particular,
the \emph{Dirac-like operators} $D_{QG}$ could be used to calculate
\emph{non-commutative} space intervals (i.e. distances) and \emph{non-commutative}
volume integrals, which, given the interpretations we made, \emph{are
the genuinely quantum distances and volumes}, since they are based
on the \emph{quantized space algebra$.\,\blacksquare$}}

\selectlanguage{spanish}%
$\;$
\selectlanguage{english}%

\subsubsection*{2.3. TQG Relational Time}

$\;$

Regarding time, we have the following situation. If we recall the
definition of the variables used in our relational construction of
the space picture (Definition 1.2), the natural generalization needed
to include time, and in this way to get a \emph{spacetime} picture,
would be 
\[
F_{f}\left(\left[h,\pi\right]\right)\doteq\int_{M}f(t,\overrightarrow{x})\,\sqrt{-g}\,\mathrm{d}^{4}x.
\]
Thus, we can see that if we want to get time into the picture, we
need to consider test functions $f$ whose support is spacetime rather
than just space. Nevertheless, there are two problems with the previous
generalization, namely: i) how do we get an identification $g_{ab}\,\longleftrightarrow\left[h,\pi\right]$
between smooth spacetime metrics $g_{ab}$ which are solutions and
phase space points $\left[h_{ab},\pi^{ab}\right]$?; ii) even if we
solve that, is the resulting assignment $f\,\longmapsto F_{f}$ injective?
Regarding point i), it's well known that the Gravitational Field Equations
have a well-posed initial value formulation, that is, an initial value
$\left[h_{ab},\pi^{ab}\right]\in X$ which satisfies the initial value
constraints, and a gauge fixing of the lapse $N$ and shift $N^{a}$
in $\varSigma$ (this if we use the so called ``wave gauge'' \cite{key-11,key-1})
determines a unique smooth and globally hyperbolic solution $g_{ab}\,\underset{N,N^{a}}{\longleftrightarrow}\left[h_{ab},\pi^{ab}\right]$.
For point ii), we have the following proposition.

$\;$

$\mathbf{Proposition\;2.4}$: For spacetime functions $f$ with support
in $[0,\epsilon)\times\varSigma$, the assignment $f\,\longmapsto F_{f}$
is injective in the limit $\epsilon\,\longrightarrow0$.

$\;$

\emph{Proof}: from now on, we take, once and for all, $N=1$ and $N^{a}=0$
(both, in $\varSigma$): thus, we will work with the identification
$g_{ab}\,\underset{1,0}{\longleftrightarrow}\left[h_{ab},\pi^{ab}\right]$.
Now, at the infinitesimal level, the infinitesimal arc of the graph
of a real variable function $h(t)\equiv\mathrm{det}\,(h_{ab})$ at
some point (say, $t=0$) can be approximated by the tangent line at
that point. To describe the latter, we need the value $h(0)$ and
its derivative $\dot{h}(0)$. In the wave gauge, $\sqrt{-g}=N\sqrt{h}$
in $\varSigma$, i.e., at $t=0$, and, with our choice $N(0)=1$,
we get $\sqrt{-g}(0)=\sqrt{h}(0).$ For $t\neq0$, 
\[
g\equiv\mathrm{det}\,(g_{ab})=-g_{00}h+\sum_{i}g_{0i}d_{i},
\]
where $d_{i}$ are the remaining determinants. In this way, 

\[
\dot{g}=-\dot{g}_{00}h-g_{00}\dot{h}+\sum_{i}\dot{g}_{0i}d_{i}+\sum_{i}g_{0i}\dot{d}_{i}.
\]
But now, at $t=0,$ $g_{00}=-1$, $g_{0i}=0,$ and then $d_{i}=0$.
Also, $\dot{h}=-2hK$ (where $K=-\frac{1}{2}h^{ab}\dot{h}_{ab}$;
note that $\pi=2\sqrt{h}K$), while it can be determined from the
wave gauge conditions that $\dot{g}_{00}=-K$ (recall that the wave
gauge condition on the coordinates is 
\[
\sum_{\mu,\nu}g^{\mu\nu}\varGamma_{\mu\nu}^{\alpha}=0,\,\forall\alpha,
\]
which means that the coordinates satisfy the wave equation; in our
variables, at $t=0$, we get from it: 
\[
\dot{g}_{00}+Nh^{ab}K_{ab}=0.)
\]
Thus, we get $\dot{g}(0)=-h(0)K(0)$. What all this means is that
the tangent line to the graph of $\sqrt{-g}$ at $t=0$ only depends
on $\left[h_{ab},\pi^{ab}\right]\in X$. Of course, due to the second
order character of the Gravitational Field Equations, the initial
data $\left[h_{ab},\pi^{ab}\right]\in X$ is arbitrary, which means
that the tangent line can be arbitrary. Note that this is not trivial
and we got it thanks to the wave gauge condition: if this condition
implied, instead, that $\dot{g}_{00}=-2K$, then we would get $\dot{g}(0)=0$,
which means that this derivative cannot be arbitrary, even if $\left[h_{ab},\pi^{ab}\right]\in X$
is. Of course, it's actually this derivative the most important thing
that we need here in order to probe beyond $\varSigma$ in the time
direction$.\,\square$

$\;$

$\mathbf{Example\;2.3}$: Let's analyse first these type of spacetime
variables in the case of standard QFT on a fixed curved spacetime
($M,g_{ab})$ \cite{key-1,key-5}. Consider the usual scalar field
$\varphi$ with a linear wave equation

\[
g_{ab}\nabla^{a}\nabla^{b}\varphi=0.
\]
Since this equation has a well-posed initial value formulation, one
can identify the phase space with the space $\mathrm{Sol}.$ of smooth
solutions. Then, for a spacetime test function $f\in C_{0}^{\infty}(M)$,
one defines the following variable on that phase space:

\[
\underline{\varphi}(f)(\varphi)\doteq\int_{M}\varphi f\boldsymbol{\epsilon}(g).
\]
The Poisson brackets for these variables are then given by

\[
\left\{ \underline{\varphi}(f),\underline{\varphi}(f')\right\} =-\Omega(Ef,Ef')
\]

\[
=-\int_{M}fEf'\boldsymbol{\epsilon}(g)=\int_{M}f'Ef\boldsymbol{\epsilon}(g),
\]
where $\Omega$ is the symplectic form on $\mathrm{Sol}.$ and the
linear map $E:C_{0}^{\infty}(M)\,\longrightarrow\mathrm{Sol}.,$$\,f\,\longmapsto Ef$
is the advanced minus the retarded solution of the wave equation with
source $f\in C_{0}^{\infty}(M)$. Now, consider the case in which
the supports of $f$ and $f'$ are \emph{causally disconnected}, i.e.,
if $f(x)f'(y)\neq0\,\Longrightarrow y\notin J^{+}(x)\bigcup J^{-}(x)$
(where $J^{+/-}$ are the causal future and past, respectively.) But
we also have that:

\[
\int_{M}f'Ef\boldsymbol{\epsilon}(g)=\int_{\mathscr{R}_{1}}f'(Af)\boldsymbol{\epsilon}(g)-\int_{\mathscr{R}_{2}}f'(Rf)\boldsymbol{\epsilon}(g),
\]
where $\mathscr{R}_{1}=supp\left(f'\right)\bigcap J^{-}\left[supp\left(f\right)\right]$
and $\mathscr{R}_{2}=supp\left(f'\right)\bigcap J^{+}\left[supp\left(f\right)\right]$,
since $A(f)$ and $R(f)$ are, respectively, the advanced and retarded
solutions. Thus, if the supports of $f$ and $f'$ are causally disconnected,
then, because of well-posedness of the field equation, clearly we
have that $\mathscr{R}_{1}=\mathscr{R}_{2}=\textrm{Ø}$ and in this
way we get

\[
\left\{ \underline{\varphi}(f),\underline{\varphi}(f')\right\} =-\Omega(Ef,Ef')=0
\]
for this case. On the other hand, if the supports are \emph{causally
connected}, we get

\[
\left\{ \underline{\varphi}(f),\underline{\varphi}(f')\right\} =-\Omega(Ef,Ef')\neq0.\,\blacksquare
\]

$\;$

With the above in mind, let's now go back to GR. In this theory, the
field equations do not have the simple form of the previous linear
wave equations, but comprise (in the so-called ``wave gauge''),
instead, what's known as a \emph{quasi-linear} system. For a scalar
field with a quasi-linear field equation, the latter takes the form

\[
g_{ab}(x;\varphi;\nabla^{c}\varphi)\nabla^{a}\nabla^{b}\varphi=F(x;\varphi;\nabla^{c}\varphi),
\]
where $g_{ab}$ is a smooth lorentzian metric and $F$ a smooth function.
These type of equations still have a well-posed initial value formulation,
but they are quite different to the standard, ordinary wave equation.
Indeed, the lorentzian metric that defines the character of the principal
symbol of the differential operator now depends on the field variable.
Of course, this makes the field equations non-linear. But, since the
principal symbol of the operator is what determines how causal influences
on this field propagate, it also means that the very causal structure
is tied now to the variation of the field variable. Unfortunately,
this fact makes all of the previous analysis of QFT in fixed curved
spacetimes to be inapplicable here. Indeed, even to this date, the
general form of the Poisson bracket for functionals like $F_{f}$
is \emph{not} known \cite{key-7}\footnote{Nevertheless, it's shown in this reference that we still get, for
the quasi-linear case, $\left\{ \underline{\varphi}(f),\underline{\varphi}(f')\right\} =0$
if the supports are causally disconnected and, presumably, $\left\{ \underline{\varphi}(f),\underline{\varphi}(f')\right\} \neq0$
if the supports are causally connected. Thus, in the quantization
below, all stages in which only this property of the brackets is used
can be considered also applicable to the general case, while the ones
in which we use the explicit form $\left\{ \underline{\varphi}(f),\underline{\varphi}(f')\right\} =-\Omega(Ef,Ef')$,
\emph{from the linear case}, are approximative.}. Thus, a complete and rigorous quantization of time by following
the recipe of our appraoch is definitely out of reach for now. Thus,
it's at this point when we start to make \emph{several} approximations,
which, we warn, may or may not be ultimately valid. We will judge
that by the reasonability of their nature and of the results that
follow from assuming them.

$\;$

For simplicity, from now on we pretend a metric solution $g_{ab}$
is just a scalar field $\varphi$. These details will not matter here
since the present analysis is only approximate and merely structural.

$\;$

$\mathbf{Definition\;2.14}$: Consider a compact curve\footnote{Actually, we take $\gamma$ as a narrow timelike/spacelike spacetime
cylinder centered at the original curve, in order to avoid a cluttering
of delta functions on the integrals.} (timelike or spacelike) segment $\gamma$ in $M$. Then, on the phase
space $X$ of GR, we define the following variables:

\[
\underline{\varphi}(f_{\gamma})(\varphi)\doteq\int_{M}\varphi f_{\gamma}\boldsymbol{\epsilon}(\varphi),\,supp\,(f_{\gamma})=\gamma,\,f_{\gamma}\in C_{0}^{\infty}(M).\,\blacksquare
\]

$\;$

$\mathbf{Proposition\;2.5}$: The previous variables in Definition
2.14 \emph{can} indeed be used to obtain $C^{\infty}(\gamma)$ via
$R_{\varphi}$ (in the limit $\epsilon\,\longrightarrow0$.)

$\;$

\emph{Proof}: as in Proposition 2.1 \cite{key-12}, we define the
following product

\[
(\underline{\varphi}(f_{\gamma}^{1})\cdot_{\gamma}\underline{\varphi}(f_{\gamma}^{2}))(\varphi)\doteq\underline{\varphi}(f_{\gamma}^{1}f_{\gamma}^{2})(\varphi),\,\forall\varphi\in X
\]

\[
=\int_{M}\varphi f_{\gamma}^{1}f_{\gamma}^{2}\boldsymbol{\epsilon}(\varphi).
\]
In the limit $\epsilon\,\longrightarrow0$, where the assignment $f_{\gamma}\,\longmapsto\underline{\varphi}(f_{\gamma})$
is injective (the term $\varphi$ in the integral can be handled by
analogous arguments as those in Proposition 2.1), we get an algebra
$\left\{ \underline{\varphi}(f_{\gamma}k_{0})\right\} _{f_{\gamma}\in C_{0}^{\infty}(M);\cdot_{\gamma}}$,
for each $k_{0}\in C^{\infty}(\gamma)$ ($k_{0}$ is non-zero for
each point in $\gamma$; all algebras for different $k_{0}$ are isomorphic
and equivalent to $C^{\infty}(\gamma)$), which, since the solution
$\varphi$ such that $k_{0}=\varphi^{-1}$ is assumed to be non-vanishing
on $\gamma$, is then bijectively mapped to $C^{\infty}(\gamma)$
via $R_{\varphi}(\underline{\varphi}(f_{\gamma}k_{0}))=f_{\gamma}$;
but, since all the algebras are isomorphic, this means that the relational
representation indeed induces the algebra $C^{\infty}(\gamma)$ of
the curve $\gamma$ from a \emph{single} algebra of phase space functions
for \emph{any} of those solutions $\varphi$$.\,\square$

$\;$

Consider some \emph{subset} of $\mathrm{Sol}.$ in which all metrics
have \emph{conformally-equivalent} causal structures (besides this
requirement, there's freedom for choosing any such subset.) Thus,
while we have many different metrics, at least we have a \emph{single}
causal structure $\mathcal{C}$. Thus, the only variation we actually
have now is that of the \emph{conformal factor} among the different
metrics. It's \emph{this} variation the one we will try to quantize.
Nevertheless, the equations are still non-linear in that factor. For
this, we will consider only the linear part for the quantization;
that is, the result will be only kinematical, since the non-linear
coupling is being ignored. In this way, what we are doing is to consider
quantum spacetime geometries which are ``benign'' and ``semi-classical''
enough such that they have a defined causal structure and which remains
more or less the same among the different geometries. Furthermore,
they are also such that the conformal factor doesn't vary non-linearly
among them. Thus, the geometries are almost classical, the only variation
is a ``residual'' one in the conformal factor, which is linear,
and hyperbolic with respect to the causal structure $\mathcal{C}$.
It's only this variation that is quantized below. 

$\;$

$\mathbf{Definition\;2.15}$: A \emph{causal set }(see, e.g., \cite{key-8}
and references therein) is a partially ordered set $\left(C,\leq_{\mathcal{C}}\right)$
such that:
\begin{lyxlist}{00.00.0000}
\item [{i)}] for all $x\in C,$ $x\leq_{\mathcal{C}}x$ (Reflexive);
\item [{ii)}] for all $x,y\in C,$ if $x\leq_{\mathcal{C}}y$ and $y\leq_{\mathcal{C}}x$,
then $x=y$ (Antisymmetric);
\item [{iii)}] for all $x,y,z\in C$, if $x\leq_{\mathcal{C}}y$ and $y\leq_{\mathcal{C}}z$,
then $x\leq_{\mathcal{C}}z$ (Transitive);
\item [{iv)}] for all $x,z\in C$, $\mid\left\{ y\in C,\,x\leq_{\mathcal{C}}y\leq_{\mathcal{C}}z\right\} \mid<\aleph_{0}$
(Locally finite)$.\,\blacksquare$
\end{lyxlist}
$\;$

$\mathbf{Definition\;2.16}$: Let $\leq_{\mathcal{C}}$ be the \emph{causal
partial order} on the points in $M$ induced by the causal structure
$\mathcal{C}$. Now, consider a graph defined on \emph{all} of the
\emph{spacetime} manifold $M$; furthermore, on each $\varSigma_{t}$,
it gives one of the graphs considered earlier in the discussion of
area, while the nodes of each of these graphs for \emph{different}
$\varSigma_{t}$ and $\varSigma_{t'}$ are connected by timelike/causal
(with respect to $\leq_{\mathcal{C}}$) edges (e.g., the previous
curve $\gamma$ could be one of them), while the edges of the graphs
in each $\varSigma_{t}$ are seen as spacelike separated (i.e., the
nodes are not related by $\leq_{\mathcal{C}}$.) We denote each node
as $P_{i}$, for a countable\footnote{Note that this is not a capricious, ad-hoc discretization, even when
we are indeed puting it by hand, since it's justified by the physical
and mathematical arguments given in all previous sections at various
points of the discussion.} index $i$. Thus, the collection of \emph{elements} (it's more convenient
to call them by this term rather than points)

\[
C\doteq\left\{ P_{i}\right\} _{i\in\mathbb{N}}^{\leq_{\mathcal{C}}}
\]
forms a causal set. Consider the edges $\mathrm{e}_{P_{N_{1}}}$ and
$\mathrm{e}_{P_{N_{2}}}$ such that $\mathrm{e}_{P_{N_{2}}}\subset J^{+}(\mathrm{e}_{P_{N_{1}}})$.
We omit the endpoints and (semi) characterize $\mathrm{e}_{P_{*}}$
simply by its initial point $P_{*}$ (that is, $\mathrm{e}_{P_{*}}$
represents \emph{any} of the edges with initial point $P_{*}$.) We
take the variables $\underline{\varphi}(\chi_{\mathrm{e}_{P_{*}}})$
(for which $supp\left(\chi_{\mathrm{e}_{P_{*}}}\right)=\mathrm{e}_{P_{*}}$),
where $\chi_{\mathrm{e}_{P_{*}}}$ is the characteristic function
of the set $\mathrm{e}_{P_{*}}$, and the algebra $\mathcal{A}_{TQG}$
that they linearly generate (with real coefficients), and promote
the labelings to elements $P_{*}$ from an \emph{abstract}, countably
infinite causal set $C$$.\,\blacksquare$ 

$\;$

$\mathbf{Lemma\;2.6}$:\foreignlanguage{spanish}{ The variables of
the proof of Propostion 2.5, $\underline{\varphi}(f_{\gamma}k_{0})$
(considering the set of \emph{all} the elements from \emph{all} the
algebras $\left\{ \underline{\varphi}(f_{\gamma}k_{0})\right\} _{f_{\gamma}\in C_{0}^{\infty}(M);\cdot_{\gamma}}$
that were defined), subjected to the process of }Definition 2.2\foreignlanguage{spanish}{,
result in the variables of Definition 2.16 (the proof is identical
to that of Lemma 2.3 \cite{key-12})$.\,\square$}

$\;$

Thus, in line with the principle of striping the continuum, this is
equivalent to replacing $f_{\mathrm{e}_{P_{*}}}$ by $k\chi_{\mathrm{e}_{P_{*}}},$
where $\chi_{\mathrm{e}_{P_{*}}}$ is the characteristic function
of the set $\mathrm{e}_{P_{*}}$ in $M$ and $k\in\mathbb{R}$. Note
how this makes the limit $\epsilon\,\longrightarrow0$ unnecessary
since $\chi_{\mathrm{e}_{P_{*}}}$ doesn't change its value along
$t$, and then these type of functions on the finite segment $\mathrm{e}_{P_{*}}$
can be taken as the ``true general functions on the infinitesimal
curve segment $\mathrm{d}\mathrm{e}_{P_{*}}$.'' Of course, $\chi_{\mathrm{e}_{P_{*}}}$
is not a smooth function, but this will not be relevant. That is,
associated to $P_{*}$, we get a $1-$dimensional abstract vector
space 
\[
\mathcal{A}_{P_{*}}\doteq\left\{ k\underline{\varphi}(P_{*})\right\} _{k\in\mathbb{R}}\cong\mathbb{R}
\]
generated by the sole element $\underline{\varphi}(P_{*})$. Now,
of course, if we quantize this we get just a commutative algebra,
since $\left\{ \underline{\varphi}(f),\underline{\varphi}(f)\right\} =-\Omega(Ef,Ef)=0$
(because $\Omega$ is a symplectic form.)

$\;$

Nevertheless, we cannot consider $\mathcal{A}_{P_{*}}$ alone, because 

\[
\mathrm{e}_{P_{N_{2}}}\bigcap J^{+}(\mathrm{e}_{P_{N_{1}}})\neq\textrm{Ø}\;\mathrm{and}\;\left\{ \underline{\varphi}(f_{\mathrm{e}_{P_{N_{1}}}}),\underline{\varphi}(f_{\mathrm{e}_{P_{N_{2}}}})\right\} =-\Omega(Ef_{\mathrm{e}_{P_{N_{1}}}},Ef_{\mathrm{e}_{P_{N_{2}}}})\neq0.
\]
But now, by the linearity of both $\Omega$ and $E$, and remembering
that we replace $f_{\mathrm{e}_{P_{*}}}$ by $k\chi_{\mathrm{e}_{P_{*}}}$,
and that $\mathcal{A}_{P_{*}}\cong\mathbb{\mathbb{R}}$, then it's
clear that, when passing to the abstract labelings, the symplectic
form $\Omega$ becomes simply a symplectic form $\widetilde{\Omega}$
on the $2-$dimensional space $\mathbb{\mathbb{R}}^{2}$ and characterized
by a simple $2\times2$ skew matrix of the form

\[
\left(\begin{array}{cc}
0 & \theta(P_{N_{1}},P_{N_{2}})\\
-\theta(P_{N_{1}},P_{N_{2}}) & 0
\end{array}\right),
\]
where

\[
\theta(P_{N_{1}},P_{N_{2}})\doteq\Omega(E\chi_{\mathrm{e}_{P_{N_{1}}}},E\chi_{\mathrm{e}_{P_{N_{2}}}}),
\]
thing which greatly simplifies the problem at the mathematical level. 

$\;$

$\mathbf{Definition\;2.17}$: \foreignlanguage{spanish}{We define:
$\mathcal{F}_{Dis.}\equiv\mathcal{A}_{ST.}\equiv\mathcal{A}_{P_{N_{1}}P_{N_{2}}}\subset\mathcal{A}_{TQG}$,
generated by the $\underline{\varphi}(f_{\mathrm{e}_{P_{N_{1}}}}),$
$\underline{\varphi}(f_{\mathrm{e}_{P_{N_{2}}}})$. Now, following
Definition 2.2, the quantum algebra $\widehat{\mathcal{A}}_{ST.}\equiv\mathcal{\widehat{A}}_{P_{N_{1}}P_{N_{2}}}$}
will be the \emph{exponentiated} (recall that the passing from the
Heisenberg to the Weyl algebra is mandatory for quantization in QFT
\cite{key-1}) \emph{quantum} algebra for the two quantized edges
$\widehat{\mathrm{e}}_{P_{N_{1}}}$ and $\widehat{\mathrm{e}}_{P_{N_{2}}}$,
and generated, then, by two abstract elements, $u_{1}$ and $u_{2}$,
which satisfy the relation

\[
u_{1}u_{2}=e^{2\pi i\theta(P_{N_{1}},P_{N_{2}})}u_{2}u_{1}.\,\blacksquare
\]

$\;$

$\mathbf{Example\;2.4}$: This algebra is well known in NCG \cite{key-3},
and is called the NC $2-$Torus\footnote{Hence the name TQG for this proposal, i.e., \emph{Toroidal Quantum
Gravity}. ``\emph{Toral}'' can be used, too.}, $\mathbb{T}_{\theta}^{2}$, with deformation parameter $\theta$,
since it corresponds to the deformation of the algebra of the classical
Torus $\mathbb{T}^{2}$ (defined as the set $\left[0,1\right]^{2}$
with opposite sides of the square identified, or as $\mathbb{T}^{1}\times\mathbb{T}^{1}$,
where $\mathbb{T}^{1}$ is the unit circle.) More precisely, we consider
the universal $C^{*}-$algebra $A_{\theta(P_{N_{1}},P_{N_{2}})}$
defined by these generators. For $(r_{1},r_{2})=r\in\mathbb{Z}^{2}$,
define the polynomials $u^{r}\doteq e^{\pi ir_{1}r_{2}\theta(P_{N_{1}},P_{N_{2}})}u_{1}^{r_{1}}u_{2}^{r_{2}}$,
then $A_{\theta(P_{N_{1}},P_{N_{2}})}\ni a=\sum_{r\in\mathbb{Z}^{2}}a_{r}u^{r}$,
where $a_{r}\in\mathbb{C}$ and the sum converges in the $C^{*}$
norm $\parallel\cdot\parallel_{C^{*}}$. The NC $2-$Torus is the
\emph{dense}, \emph{smooth} $*-$subalgebra, pre$C^{*}-$algebra $C^{\infty}(\mathbb{T}_{\theta}^{2})=\mathcal{\widehat{A}}_{\theta(P_{N_{1}},P_{N_{2}})}$
of $A_{\theta(P_{N_{1}},P_{N_{2}})}$, that is, $r\,\longmapsto a_{r}$
is of \emph{rapid decay} (the $C^{*}-$algebra $A_{\theta(P_{N_{1}},P_{N_{2}})}$,
of course, gives the algebra of \emph{continuous} NC-``functions''.)
Now, $\mathbb{T}^{2}$ forms a Lie group, and we can define an action
of this group on $A_{\theta(P_{N_{1}},P_{N_{2}})}$ by setting, for
any $z\in\mathbb{T}^{2}$,

\[
z\cdot u^{r}\doteq z_{1}^{r_{1}}z_{2}^{r_{2}}u^{r}.
\]
This action is generated by the following two, commuting derivations:

\[
\delta_{j}(a)\doteq\left[\frac{\mathrm{d}}{\mathrm{d}t}e^{2\pi i\phi_{j}}\cdot a\right]\mid_{t=0},\,j=1,2,
\]
where $e^{2\pi i\phi_{j}}$ belongs to the circle group $\mathbb{T}^{1}$
(i.e., complex numbers of norm equal to $1$.) The following trace
defines a faithfull algebraic state on $A_{\theta(P_{N_{1}},P_{N_{2}})}$:

\[
\omega(a)\doteq a_{0}.
\]
Note that

\[
\omega(a^{*}a)=\sum_{r\in\mathbb{Z}^{2}}\mid a_{r}\mid^{2}\leq\parallel a\parallel_{C^{*}}^{2}.
\]
With it, we can form the GNS representation $\pi_{\omega}$ \cite{key-5}
of the algebra (that is, $\left(a,b\right)_{\mathcal{H}_{\omega}}\doteq\omega(a^{*}b)=\left(a^{*}b\right)_{0}=\sum_{r\in\mathbb{Z}^{2}}a_{r}^{*}b_{r}$.)
We denote as $\underline{a}$ to the algebra elements when seen as
elements in the Hilbert space $\mathcal{H}_{\omega}$ of the representation
(whose cyclic vector is $\underline{I}$ and is irreducible) and the
same for the derivations when seen as acting on the algebra representation.
We can now define a Dirac operator by setting (on the dense domain
$\underline{\mathcal{\widehat{A}}}_{\theta(P_{N_{1}},P_{N_{2}})}$)

\[
D_{P_{N_{1}}P_{N_{2}}}\doteq-i\mathrm{a}(P_{N_{1}},P_{N_{2}})^{-\frac{1}{2}}\left[\sigma_{1}\underline{\delta}_{1}+\sigma_{2}\underline{\delta}_{2}\right],
\]
where $\sigma_{1},\sigma_{2}$ are the first two Pauli matrices. In
this way, one can see that 
\[
\left(\pi_{\omega}\left[A_{\theta(P_{N_{1}},P_{N_{2}})}\right],\mathcal{H}_{\omega},D_{P_{N_{1}}P_{N_{2}}}\right)
\]
forms a $2-$dimensional spectral triple (the eigenbasis of $D_{P_{N_{1}}P_{N_{2}}}^{2}$
is given by $\left\{ \underline{u}^{r}\right\} _{r\in\mathbb{Z}^{2}}$
tensored with the canonical basis of $\mathbb{C}^{2}$, with eigenvalues
$\lambda_{r}=\mathrm{a}(P_{N_{1}},P_{N_{2}})^{-1}4\pi^{2}r\cdot r$
of multiplicity $M_{r}=2$.) Furthermore (after some adequate normalization),

\[
\mathrm{area}_{D_{P_{N_{1}}P_{N_{2}}}}(\widehat{\mathcal{R}}_{N_{1}N_{2}})\doteq\mathrm{tr}^{+}(\mid D_{P_{N_{1}}P_{N_{2}}}\mid^{-2})
\]

\[
=\mathrm{a}(P_{N_{1}},P_{N_{2}}).
\]
See \cite{key-3} for more details$.\,\blacksquare$

$\;$

$\mathbf{Remark\;2.10}$: Note that this \emph{doesn't} mean that
the geometry of spacetime at the quantum level is simply a NC Torus,
since the physical interpretation of the algebra elements that we
made before doesn't lead to this. Indeed, for $\theta=0$, $u_{1}$
and $u_{2}$ generate a classical, continuum Torus, while the $\underline{\varphi}(\chi_{\mathrm{e}_{P_{N_{1}}}})$
and $\underline{\varphi}(\chi_{\mathrm{e}_{P_{N_{2}}}})$ generate
what's left of the algebra of the edges \emph{after we eliminated
the continuum}. Thus, the Torus geometrical interpretation from NCG
will not be relevant here, and we are just using/borrowing its algebra
for our own purposes and particular interpretations$.\,\blacksquare$

$\;$

$\mathbf{Remark\;2.11}$: The physical interpretation of $\mathrm{area}_{D_{P_{N_{1}}P_{N_{2}}}}(\widehat{\mathcal{R}}_{N_{1}N_{2}})$
\emph{when} $P_{N_{2}}$ \emph{is the endpoint} of $\mathrm{e}_{P_{N_{1}}}$
is\footnote{We also take $\mathrm{e}_{P_{N_{1}}}$ as \emph{future-directed (fd)
timelike} and $\mathrm{e}_{P_{N_{2}}}$ as \emph{spacelike} (with
respect to the causal partial order.)} as the quantized riemannian area of the coordinate surface $\mathcal{R}_{N_{1}N_{2}}$
of $\mathrm{e}_{P_{N_{1}}}\times\mathrm{e}_{P_{N_{2}}}$ in coordinate
space (that is, we parallelly propagate $\mathrm{e}_{P_{N_{1}}}$
along $\mathrm{e}_{P_{N_{2}}}$, then set $P_{N_{2}}$ as having coordinates
$(x_{1},x_{2})=(1,0)$ and the endpoint of $\mathrm{e}_{P_{N_{2}}}$
as $(1,1)$, and then map the whole surface to a rectangle $\mathcal{R}_{N_{1}N_{2}}$
in coordinate space). Of course, we lose the lorentzian character,
but this is not a problem since that is already being taken into account
by the causal structure $\mathcal{C}$, but this riemannian metric
still gives information about the duration and lenght of the \emph{finite
$2-$dimensional process} described by the spacetime surface $\mathrm{e}_{P_{N_{1}}}\times\mathrm{e}_{P_{N_{2}}}$,
that is, a process which happens to the whole edge $\mathrm{e}_{P_{N_{2}}}$
for the duration of $\mathrm{e}_{P_{N_{1}}}$. In this way, given
that, in order to take into account the causal relation between the
two edges, we need to consider the combined quantized algebras of
each one (which in turns gives a noncommutative algebra due to the
causal relation) and that the nontrivial\footnote{The $1-$dimensional Dirac operators of each edge (which give the
usual classical metric to curves) lifted to the full algebra do not
count since their action is trivial on the part corresponding to the
algebra of the other edge.} Dirac operator on it forms a triple of dimension $2$, then this
leads us to take the (somewhat expected) view in which

$\;$

\emph{spacetime events at a quantum level (which we denote as $\mathsf{e}_{N_{1}N_{2}}$,
for the one corresponding to $P_{N_{1}},P_{N_{2}}$) are given by
$2-$dimensional processes, i.e., processes that happen to a thing
of finite length in a finite amount of proper time; furthermore, for
a given quantum spacetime (in the sense defined here), the passing
from one event to another is evidently discretized}\footnote{This applies only to individual quantum spacetimes. The passing from
one to another may be given by a continuous change in $\mathrm{a}(P_{N_{1}},P_{N_{2}})$
(the points fixed.) Nevertheless, since in LQG the values of length
are discretized, this probably means that $\mathrm{a}(P_{N_{1}},P_{N_{2}})$
is just the product of two different discrete variables, and then
the mentioned change would also be discrete because of this.}\emph{ since the graph is countably infinite}$.\,\blacksquare$

$\text{\;}$

$\mathbf{Lemma\;2.7}$: There exists a state $\underline{\psi}_{N_{1}N_{2}}\in\mathcal{H}_{\omega}$
such that

\[
\left(\underline{\psi}_{N_{1}N_{2}},D_{P_{N_{1}}P_{N_{2}}}^{2}\underline{\psi}_{N_{1}N_{2}}\right)_{\mathcal{H}_{\omega}}=\mathrm{a}(P_{N_{1}},P_{N_{2}})^{-1}=\mathrm{area}_{D_{P_{N_{1}}P_{N_{2}}}}(\widehat{\mathcal{R}}_{N_{1}N_{2}})^{-1}.
\]

$\;$

\emph{Proof}: since the triple is $2-$dimensional, then $\sum_{k=1}^{N}M_{k}\lambda_{k}(\mid D_{P_{N_{1}}P_{N_{2}}}\mid^{-2})$
must diverge as $\sum_{k=1}^{N}\frac{1}{k}\sim\mathrm{ln}\,N$ when
$N\,\longrightarrow\infty$. This is similar to the inverse of the
harmonic oscillator Hamiltonian in standard QM, for which

\[
\psi_{s}(x)\doteq\pi^{-\frac{1}{4}}e^{-(x-s)^{2}/2},\,s\in\mathbb{R},
\]
is a coherent/Gaussian state. This state can be written in terms of
the eigenbasis $\left\{ \varphi_{k}\right\} _{k\in\mathbb{N}\bigcup\left\{ 0\right\} }$
of the Hamiltonian as

\[
\psi_{s}(x)=e^{-s^{2}/2}\sum_{k=0}^{\infty}\sqrt{\frac{\left(s^{2}/2\right)^{k}}{k!}}\varphi_{k}(x).
\]
This suggests to define the following state in $\mathcal{H}_{\omega}$:

\[
\underline{\psi}_{s}\doteq\frac{\left(e^{s^{2}/4}-1\right)^{-1}}{2}\sum_{\underset{r_{1}\neq0;r_{2}\neq0}{r\in\mathbb{Z}^{2}}}\sqrt{\frac{\left(s^{2}/4\right)^{\mid r_{1}\mid}}{\mid r_{1}\mid!}}\sqrt{\frac{\left(s^{2}/4\right)^{\mid r_{2}\mid}}{\mid r_{2}\mid!}}\underline{u}^{r}.
\]
First we need to check if this series defines a state in the first
place. Indeed\footnote{Note that $\underline{a}_{(0,0)}^{\underline{\psi}_{s}}=(\underline{\psi}_{s},\underline{u}^{(0,0)})_{\mathcal{H}_{\omega}}=0$,
where $\underline{u}^{(0,0)}=\underline{I}$, and $\underline{a}_{(r_{1},0)}^{\underline{\psi}_{s}}=\underline{a}_{(0,r_{2})}^{\underline{\psi}_{s}}=0,\,\forall r\in\mathbb{Z}^{2}$.}:

\[
\parallel\underline{\psi}_{s}\parallel_{\mathcal{H}_{\omega}}^{2}=\frac{\left(e^{s^{2}/4}-1\right)^{-2}}{4}\sum_{\underset{r_{1}\neq0;r_{2}\neq0}{r\in\mathbb{Z}^{2}}}\frac{\left(s^{2}/4\right)^{\mid r_{1}\mid}}{\mid r_{1}\mid!}\frac{\left(s^{2}/4\right)^{\mid r_{2}\mid}}{\mid r_{2}\mid!}
\]

\[
=1,\,\forall s\in\mathbb{R}.
\]
Furthermore,

\[
\left(\underline{\psi}_{s},D_{P_{N_{1}}P_{N_{2}}}^{2}\underline{\psi}_{s}\right)_{\mathcal{H}_{\omega}}=\mathrm{a}(P_{N_{1}},P_{N_{2}})^{-1}8\pi^{2}\frac{\left(e^{s^{2}/4}-1\right)^{-2}}{4}\times
\]

\[
\sum_{\underset{r_{1}\neq0;r_{2}\neq0}{r\in\mathbb{Z}^{2}}}\frac{\left(s^{2}/4\right)^{\mid r_{1}\mid}}{\mid r_{1}\mid!}\frac{\left(s^{2}/4\right)^{\mid r_{2}\mid}}{\mid r_{2}\mid!}(r_{1}^{2}+r_{2}^{2});
\]
noting that

\[
r_{*}^{2}s^{2\mid r_{*}\mid}=\frac{s^{2}}{4}\frac{\mathrm{d}^{2}}{\mathrm{d}s^{2}}s^{2\mid r_{*}\mid},
\]
then

\[
\sum_{\underset{r_{1}\neq0;r_{2}\neq0}{r\in\mathbb{Z}^{2}}}\frac{\left(s^{2}/4\right)^{\mid r_{1}\mid}}{\mid r_{1}\mid!}\frac{\left(s^{2}/4\right)^{\mid r_{2}\mid}}{\mid r_{2}\mid!}(r_{1}^{2}+r_{2}^{2})
\]

\[
=\frac{s^{2}}{2}\sum_{\underset{r_{1}\neq0;r_{2}\neq0}{r\in\mathbb{Z}^{2}}}\left(\frac{\mathrm{d}^{2}}{\mathrm{d}s^{2}}\frac{\left(s^{2}/4\right)^{\mid r_{1}\mid}}{\mid r_{1}\mid!}\right)\left(\frac{\left(s^{2}/4\right)^{\mid r_{2}\mid}}{\mid r_{2}\mid!}\right)
\]

\[
=e^{s^{2}/4}s^{2}\left(\frac{s^{2}}{2}+1\right)\left(e^{s^{2}/4}-1\right);
\]
in this way:

\[
\left(\underline{\psi}_{s},D_{P_{N_{1}}P_{N_{2}}}^{2}\underline{\psi}_{s}\right)_{\mathcal{H}_{\omega}}=\mathrm{a}(P_{N_{1}},P_{N_{2}})^{-1}\pi^{2}\frac{\left(e^{s^{2}/4}-1\right)^{-1}}{2}\times
\]

\[
e^{s^{2}/4}s^{2}\left(\frac{s^{2}}{2}+1\right).
\]
Now, consider the following equation:

\[
\pi^{2}\frac{\left(e^{s^{2}/4}-1\right)^{-1}}{2}e^{s^{2}/4}s^{2}\left(\frac{s^{2}}{2}+1\right)=1.
\]
It has a \emph{single} solution, given by the \emph{unique} limit
$s\longrightarrow0$. We denote $\underline{\psi}_{s\longrightarrow0}$
as $\underline{\psi}_{N_{1}N_{2}}$. Thus\footnote{In the \emph{dual} space, we get $\left(\underline{\psi^{*}}_{N_{1}N_{2}},D_{P_{N_{1}}P_{N_{2}}}^{-2}\underline{\psi^{*}}_{N_{1}N_{2}}\right)_{\mathcal{H}_{\omega}^{*}}=\mathrm{a}(P_{N_{1}},P_{N_{2}})$,
where the limit $s\longrightarrow0$ \emph{do} commutes with $D_{P_{N_{1}}P_{N_{2}}}^{-2}$
because the latter is compact, and therefore bounded (i.e., continuous.)}:

\[
\left(\underline{\psi}_{N_{1}N_{2}},D_{P_{N_{1}}P_{N_{2}}}^{2}\underline{\psi}_{N_{1}N_{2}}\right)_{\mathcal{H}_{\omega}}=\mathrm{a}(P_{N_{1}},P_{N_{2}})^{-1}=\mathrm{area}_{D_{P_{N_{1}}P_{N_{2}}}}(\widehat{\mathcal{R}}_{N_{1}N_{2}})^{-1}.\,\square
\]

$\;$

$\mathbf{Remark\;2.12}$: In this way, at least formally\footnote{For operators $\pi_{\omega}(a)$ either in the continuous or smooth
algebras, then the expression $\mathcal{F}_{a}(\omega_{\underline{\psi}_{N_{1}N_{2}}})\doteq\omega_{\underline{\psi}_{N_{1}N_{2}}}(a)\doteq\left(\underline{\psi}_{N_{1}N_{2}},\pi_{\omega}(a)\underline{\psi}_{N_{1}N_{2}}\right)_{\mathcal{H}_{\omega}}$
\emph{does} define a genuine pure algebraic state (recall that pure
states are in $1-1$ correspondence with the points $x$ of the spacetime
$M$ in the classical case and $\mathcal{F}_{f}(\omega_{x})\doteq\omega_{x}(f)=f(x)$
is just the Gelfand transform \cite{key-5}, where $\mathcal{F}_{f}$
corresponds to the continous function $f$ on $M$.)}, we can see $\underline{\psi}_{N_{1}N_{2}}$ as a ``pure algebraic
state'' $\omega_{\underline{\psi}_{N_{1}N_{2}}}$ which acts as 
\[
\omega_{\underline{\psi}_{N_{1}N_{2}}}(D_{P_{N_{1}}P_{N_{2}}}^{2})\doteq\left(\underline{\psi}_{N_{1}N_{2}},D_{P_{N_{1}}P_{N_{2}}}^{2}\underline{\psi}_{N_{1}N_{2}}\right)_{\mathcal{H}_{\omega}}.
\]
Therefore, we can set:

\[
\mathcal{F}_{D_{P_{N_{1}}P_{N_{2}}}^{2}}(\omega_{\underline{\psi}_{N_{1}N_{2}}})\doteq\omega_{\underline{\psi}_{N_{1}N_{2}}}(D_{P_{N_{1}}P_{N_{2}}}^{2})=\mathrm{area}_{D_{P_{N_{1}}P_{N_{2}}}}(\widehat{\mathcal{R}}_{N_{1}N_{2}})^{-1}.
\]
We can interpret this in the sense that $\mathcal{F}_{D_{P_{N_{1}}P_{N_{2}}}^{2}}$
represents the inverse area ``function'' and that there's just a
single point $\omega_{\underline{\psi}_{N_{1}N_{2}}}$ in which its
value is defined. After we introduce more events (and when they \emph{commute}
with each other), we can lift the different $\mathcal{F}_{D_{P_{N_{1}}P_{N_{2}}}^{2}}$
into a function $\mathcal{F}_{\mathrm{area}_{N_{1}N_{2}}}$ ``on''
the algebra of \emph{all} the considered points and which can be interpreted
as the analogue of a ``\emph{characteristic function}'' for the
spacetime ``point'' $\omega_{\underline{\psi}_{N_{1}N_{2}}}$ (although,
the value of the function is $\mathrm{area}_{D_{P_{N_{1}}P_{N_{2}}}}(\widehat{\mathcal{R}}_{N_{1}N_{2}})^{-1}$
rather than just $1$, that is, the event $\mathsf{e}_{N_{1}N_{2}}$
is marked or given a physical interpretation \emph{in a relational
manner} by a property of the quantum Gravitational Field.) In the
classical case, its analogue would be\footnote{Note that $\chi_{\left\{ P\right\} }$ is also \emph{not} in the algebra
of \emph{continuous} or \emph{smooth} functions either, but is still
a function on the points $x$ in the most basic sense of the term.
Note that $\chi_{\left\{ P\right\} }(P')=\delta_{P}(\left\{ P'\right\} )$,
where $\delta_{P}$ is the Dirac measure on $M$ associated to $P$
and $\left\{ P'\right\} $ is a singleton.} $\phi_{j}(P)\chi_{\left\{ P\right\} }:M\,\longrightarrow\mathbb{R}$,
where $\phi_{j}(P)\chi_{\left\{ P\right\} }(x)=0$ if $M\ni x\neq P$,
$\phi_{j}(P)\chi_{\left\{ P\right\} }(x)=\phi_{j}(P)$ if $x=P$,
and $\phi_{j}(P)$ are coordinates based on properties of the field
(like distances and proper times)$.\,\blacksquare$

$\;$

The next step now is to add more events. To this end, consider the
previous event $\mathsf{e}_{N_{1}N_{2}}$, a fd\emph{ timelike} edge
$\overline{\mathrm{e}}_{P_{N_{2}}}$ whose \emph{initial} point is
$P_{N_{2}}$ (recall that $P_{N_{2}}$ is \emph{also} the initial
point of $\mathrm{e}_{P_{N_{2}}}$, but this edge is spacelike), and
the event $\mathsf{e}_{\overline{N}_{2}N_{3}}$ generated by $\overline{\mathrm{e}}_{P_{N_{2}}}$
and the spacelike edge $\mathrm{e}_{P_{N_{3}}}$, where $P_{N_{3}}$
is the endpoint of $\overline{\mathrm{e}}_{P_{N_{2}}}$ (for notational
consistency, from now on we will denote the previous timelike edge
$\mathrm{e}_{P_{N_{1}}}$ as $\overline{\mathrm{e}}_{P_{N_{1}}}$,
and the event $\mathsf{e}_{N_{1}N_{2}}$ as $\mathsf{e}_{\overline{N}_{1}N_{2}}$,
even when we are not considering any spacelike edge from $P_{N_{1}}$.) 

$\;$

$\mathbf{Definition\;2.18}$: For \emph{two events}, we have $4$
generators, namely, $u_{1}$, corresponding to $\overline{\mathrm{e}}_{P_{N_{1}}}$,
$u_{2}$, to $\mathrm{e}_{P_{N_{2}}}$, $u_{3}$, to $\overline{\mathrm{e}}_{P_{N_{2}}}$,
and $u_{4}$, to $\mathrm{e}_{P_{N_{3}}}$. In this way, the commutation
relations are given by

\[
u_{1}u_{2}=e^{2\pi i\theta(\overline{P}_{N_{1}},P_{N_{2}})}u_{2}u_{1},
\]

\[
u_{3}u_{4}=e^{2\pi i\theta(\overline{P}_{N_{2}},P_{N_{3}})}u_{4}u_{3},
\]

\[
u_{1}u_{3}=e^{2\pi i\theta(\overline{P}_{N_{1}},\overline{P}_{N_{2}})}u_{3}u_{1},
\]

\[
...\,etc.
\]
We can recognize the first two as the algebras $\mathcal{\widehat{A}}_{\mathsf{e}_{\overline{N}_{1}N_{2}}}$,
$\mathcal{\widehat{A}}_{\mathsf{e}_{\overline{N}_{2}N_{3}}}$, corresponding,
respectively, to the events $\mathsf{e}_{\overline{N}_{1}N_{2}}$,
$\mathsf{e}_{\overline{N}_{2}N_{3}}$; we denote the one generated
by the third relation as $\mathcal{\widehat{A}}_{\overline{P}_{N_{1}}\overline{P}_{N_{2}}}$$.\,\blacksquare$ 

$\;$

$\mathbf{Example\;2.5}$: These four generators form what is actually
known as the \emph{NC $4-$Torus} \cite{key-3}, that is, the smooth
subalgebra $C^{\infty}(\mathbb{T}_{\Theta}^{4})$ of the universal
$C^{*}-$algebra $A_{\Theta}$ defined by

\[
A_{\Theta}\ni a\doteq\sum_{r\in\mathbb{Z}^{4}}a_{r}u^{r},
\]

\[
u^{r}\doteq e^{\pi i\sum_{\underset{j<k}{j,k=1}}^{4}r_{j}\Theta_{jk}r_{k}}u_{1}^{r_{1}}u_{2}^{r_{2}}u_{3}^{r_{3}}u_{4}^{r_{4}},
\]

\[
\Theta\doteq\left(\begin{array}{cccc}
0 & \theta(\overline{P}_{N_{1}},P_{N_{2}}) & \theta(\overline{P}_{N_{1}},\overline{P}_{N_{2}}) & \theta(\overline{P}_{N_{1}},P_{N_{3}})\\
-\theta(\overline{P}_{N_{1}},P_{N_{2}}) & 0 & \theta(P_{N_{2}},\overline{P}_{N_{2}}) & \theta(P_{N_{2}},P_{N_{3}})\\
-\theta(\overline{P}_{N_{1}},\overline{P}_{N_{2}}) & -\theta(P_{N_{2}},\overline{P}_{N_{2}}) & 0 & \theta(\overline{P}_{N_{2}},P_{N_{3}})\\
-\theta(\overline{P}_{N_{1}},P_{N_{3}}) & -\theta(P_{N_{2}},P_{N_{3}}) & -\theta(\overline{P}_{N_{2}},P_{N_{3}}) & 0
\end{array}\right).
\]
Of course, the subalgebra of elements $a$ such that $a_{(r_{1},r_{2},r_{3},r_{4})}=0$,
for \emph{all} $r_{3}\neq0$, $r_{4}\neq0$, is isomorphic to $\mathcal{\widehat{A}}_{\mathsf{e}_{\overline{N}_{1}N_{2}}}$,
etc. Again, one can define the action of $z\in\mathbb{T}^{4}$ as

\[
z\cdot u^{r}\doteq z_{1}^{r_{1}}z_{2}^{r_{2}}z_{3}^{r_{3}}z_{4}^{r_{4}}u^{r},
\]
and the associated derivations

\[
\delta_{j}(a)\doteq\left[\frac{\mathrm{d}}{\mathrm{d}t}e^{2\pi i\phi_{j}}\cdot a\right]\mid_{t=0},\,j=1,2,3,4;
\]
also the algebraic state

\[
\omega(a)\doteq a_{0},
\]
its GNS representation, and, finally, the Dirac operator there\footnote{Where $\gamma_{j}$ are the generators of the action of the Clifford
algebra $\mathbb{C}\mathrm{l}(\mathbb{R}^{4})$ on $\mathbb{C}^{4}$.}:

\[
D_{\varTheta}\doteq-i\left[\gamma_{1}\underline{\delta}_{1}+\gamma_{2}\underline{\delta}_{2}+\gamma_{3}\underline{\delta}_{3}+\gamma_{4}\underline{\delta}_{4}\right].
\]
The generalization of all this to a finite number $n\in\mathbb{N}$
of events should be obvious by now. Of course, the algebra will be
given by a NC $2n-$Torus$.\,\blacksquare$ 

$\;$

$\mathbf{Remark\;2.13}$: Note that this NC $2n-$Torus \emph{doesn't}
represent quantum spacetime (to start, it would be $2n-$dimensional!),
what it actually represents is \emph{quantum phase space}. The reason
for this is that the relational representation only works for infinitesimal
(in time) events at the classical level, while the time separation
between two timelike edges fails to be of this type. Thus, the situation
is the following: the full NC $2n-$Torus quantum algebra is the algebra
of phase space, and, if we restrict to the sub-algebras corresponding
to the NC $2-$Tori that define our individual quantum events, then
we can now apply the relational representration on each of them and
obtain a piece of quantum spacetime there. Furthermore, since now
the Field portion that defines the events is \emph{finite}, then these
events can be seen as states in the quantum phase space too (more
precisely, the \emph{spacetime support}, represented by vectors like
$\underline{\psi}_{N_{1}N_{2}}$, of these states, which, in combination
with a Dirac operator, do give the state), unlike the classical case
and the infinitesimal portions there. Nevertheless, the collection
of all these geometries \emph{doesn't} form something that represents
a metric on ``the union of these regions'' (in particular, the previous
$D_{\varTheta}$ cannot be interpreted as this desired metric.) Note
that if we change the value of the area of event $\mathsf{e}_{\overline{N}_{2}N_{3}}$,
i.e., changing $D_{P_{N_{2}}P_{N_{3}}},$ then it must be considered
to be a \emph{different} event/state and its spectral triple \emph{direct
summed} (at the unexponentiated level) with the one of the previous
value in order to have a phase space describing both$.\,\blacksquare$

$\;$

We go back now to the parameter $\theta(\overline{P}_{N_{1}},P_{N_{2}})$.
We flat in space the spacetime cylinders to spacetime sheets in the
$xt$ spacetime plane and then, after integration of delta functions
on the remaining two spatial variables, we have:

\[
\theta(\overline{P}_{N_{1}},P_{N_{2}})=\int_{\mathrm{e}_{P_{N_{2}}}\bigcap J(\overline{\mathrm{e}}_{P_{N_{1}}})_{xt}}\chi_{\mathrm{e}_{P_{N_{2}}}}(x,t)(E\chi_{\overline{\mathrm{e}}_{P_{N_{1}}}})(x,t)\,\sqrt{-g}\mathrm{d}x\mathrm{d}t;
\]
now, since $\mathrm{e}_{P_{N_{2}}}\bigcap J^{-}(\overline{\mathrm{e}}_{P_{N_{1}}})_{xt}=\emptyset$,
then

\[
\theta(\overline{P}_{N_{1}},P_{N_{2}})=\int_{\mathrm{e}_{P_{N_{2}}}\bigcap J^{+}(\overline{\mathrm{e}}_{P_{N_{1}}})_{xt}}\chi_{\mathrm{e}_{P_{N_{2}}}}(x,t)(R\chi_{\overline{\mathrm{e}}_{P_{N_{1}}}})(x,t)\,\sqrt{-g}\mathrm{d}x\mathrm{d}t.
\]
Consider $\mathrm{e}_{P_{N_{2}}}$ to be of coordinate \emph{width}
$\delta$ and small coordinate \emph{time extension} $\epsilon'$,
$\overline{\mathrm{e}}_{P_{N_{1}}}$ to be of \emph{time extension}
$\epsilon\geq\epsilon'$ and small \emph{width} $\delta'\leq\delta$,
and the wave to fall to zero abruptly only when it gets close to its
edges (remaining constant and just equal to $1$ inside them). Since
that integral is \emph{only} valid at the classical level, the relevant
informtion for the quantum theory must be judiciously extracted from
it. If we use riemannian normal coordinates around $P_{N_{1}}$, then
$\sqrt{-g}\mid_{P_{N_{1}}}=1$; furthermore, in the quantum theory,
the \emph{minimal} possible non-zero length is given by the length
of $\mathrm{e}_{P_{N_{2}}}$, while the \emph{minimal} possible non-zero
time is given by the time of $\overline{\mathrm{e}}_{P_{N_{1}}}$,
and this implies $\epsilon=\epsilon'$ and $\delta'=\delta$. Thus:

\[
\theta(\overline{P}_{N_{1}},P_{N_{2}})=lim_{\epsilon,\delta\,\longrightarrow0}\int_{[0,\epsilon]}\int_{[0,\delta]}\mathrm{d}x\mathrm{d}t.
\]
Thus, at the quantum level, all this suggests to take:

$\;$

$\mathbf{Definition\;2.19}$: $\theta(\overline{P}_{N_{1}},P_{N_{2}})\doteq\mathrm{area}_{D_{\overline{P}_{N_{1}}P_{N_{2}}}}(\widehat{\mathcal{R}}_{\overline{N}_{1}N_{2}})=\mathrm{a}(\overline{P}_{N_{1}},P_{N_{2}}).\,\blacksquare$

$\;$

This choice makes the formalism self-contained, since $\mathrm{a}(\overline{P}_{N_{1}},P_{N_{2}})$
is, initially, just a parameter that enters in the definition of $D_{\overline{P}_{N_{1}}P_{N_{2}}}$.
An analogous argument also suggests to take:

$\;$

$\mathbf{Definition\;2.20}$: $\theta\doteq\theta(\overline{P}_{N_{1}},\overline{P}_{N_{2}})\doteq\mathrm{a}(\overline{P}_{N_{2}},P_{N_{3}}),$
and $\theta(\overline{P}_{N_{2}},P_{N_{3}})\doteq\mathrm{a}(\overline{P}_{N_{2}},P_{N_{3}}).\,\blacksquare$

$\;$

$\mathbf{Definition\;2.21}$: Consider now the curve $\gamma_{\overline{N}_{2}\overline{N}_{n}}\doteq\overline{\mathrm{e}}_{P_{N_{2}}}\bigcup...\bigcup\overline{\mathrm{e}}_{P_{N_{n}}}$
($n\geq2$.) Thus, by the basic properties of the integral:

\[
\theta(n)\equiv\theta(\gamma_{\overline{N}_{2}\overline{N}_{n}},P_{N_{n+1}})\doteq\sum_{j=2}^{n}\mathrm{a}(\overline{P}_{N_{j}},P_{N_{j+1}})
\]

\[
=\mathrm{a}(\overline{P}_{N_{2}},P_{N_{3}})+...+\mathrm{a}(\overline{P}_{N_{n}},P_{N_{n+1}}),
\]
and, assuming that an event $\mathsf{e}_{\gamma_{\overline{N}_{2}\overline{N}_{n}}N_{n+1}}$
with the area in the left hand side below exists\footnote{Note that this event \emph{is not} the composition of the events that
comprise the previous curve, since that's not an infinitesimal displacement
in the classical case. That is, the event in consideration is such
that just shares, numerically, the same area value as that composition.
If the area of the events is discretized as $a_{m}=ma_{0}$, where
$m\in\mathbb{N}$, then the existence of $\mathsf{e}_{\gamma_{\overline{N}_{2}\overline{N}_{n}}N_{n+1}}$
is guaranteed.} in the phase space,

\[
\mathrm{a}(\gamma_{\overline{N}_{2}\overline{N}_{n}},P_{N_{n+1}})\doteq\theta(\gamma_{\overline{N}_{2}\overline{N}_{n}},P_{N_{n+1}}).\,\blacksquare
\]

$\;$

$\mathbf{Remark\;2.14}$: What this means is that we can now consider
a process in which one goes from $\mathsf{e}_{\overline{N}_{1}N_{2}}$
to the start of $\mathsf{e}_{\overline{N}_{n+1}N_{n+2}}$, with fixed
$n\in\{2,3,...\}$ (that is, we have $n\in\mathbb{N}$ events), but
\emph{without} ``making a pause'' in what would be the ``intermediate
events'': that is, if any of the intermediate events happens, then
$\mathsf{e}_{\gamma_{\overline{N}_{2}\overline{N}_{n}}N_{n+1}}$ \emph{cannot}
happen after it, and viceversa, $\mathsf{e}_{\gamma_{\overline{N}_{2}\overline{N}_{n}}N_{n+1}}$
is as \emph{an elementary, non-compound} event. This new event now
takes the role of $\mathsf{e}_{\overline{N}_{2}N_{3}}$ in the previous
scheme with only \emph{two} ``glued'' events, but now taking $\theta\equiv\theta(n)\equiv\theta(\gamma_{\overline{N}_{2}\overline{N}_{n}},P_{N_{n+1}})$.

$\;$

\begin{figure}[H]
$\qquad\qquad\qquad\qquad\qquad\qquad$\includegraphics[scale=1.5]{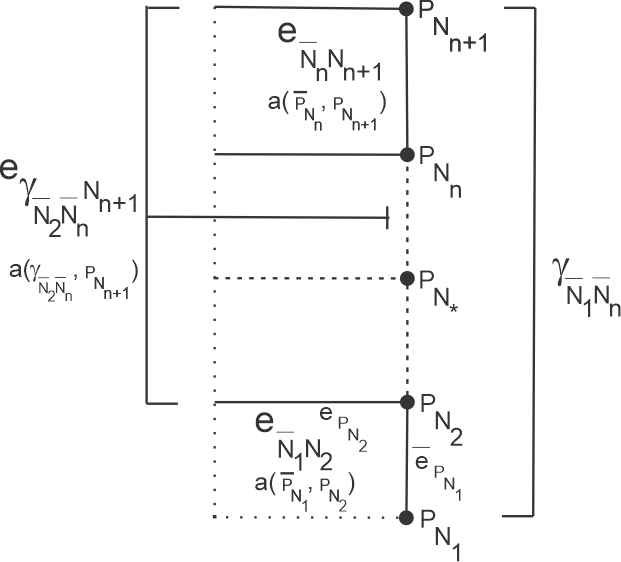}\caption{Timelike Curve in Quantum Spacetime.}
\end{figure}

In this way, as $n$ advances, we can interpret this in the sense
that the event $\mathsf{e}_{\overline{N}_{n+1}N_{n+2}}$ is more ``far
away'' from $\mathsf{e}_{\overline{N}_{1}N_{2}}$ in terms of proper
time along the quantum timelike curve defined by the succession of
the events $\mathsf{e}_{\overline{N}_{1}N_{2}}$ to $\mathsf{e}_{\overline{N}_{n+1}N_{n+2}}$.

$\;$

Now, if we \emph{experience} an \emph{elementary} process, then, since
there aren't any ``instants'' of time ``in the middle'', we would
simply age a \emph{finite} amount of time \emph{abruptly}; that is,
there's a \emph{change}, which, \emph{by itself}, doesn't introduce
any \emph{extra} proper time, the latter, instead, being fully accounted
by the metric of the event and given ``all at once'', as a photon
gives all of its finite energy to an electron in an atom all at once.
In the case of the previous elementary, ``would be compound'' events
$\mathsf{e}_{\gamma_{\overline{N}_{2}\overline{N}_{n}}N_{n+1}}$,
if there's something like a particle in some of the intermediate events,
then, if the system undergoes the elementary process $\mathsf{e}_{\gamma_{\overline{N}_{2}\overline{N}_{n}}N_{n+1}}$,
it \emph{never interacts} with the particle, it just ``tunnels it''$.\,\blacksquare$

$\;$

Before continuing, we will take $\theta(\overline{P}_{N_{1}},\overline{P}_{N_{2}})\equiv\theta$,
in order to simplify the notation.

$\;$

In the classical case, the time evolution in phase space is a map
$t\,\longmapsto\phi_{t}^{*}s_{0}$. Here we define it in the same
way, but now, since the states/events are pure states in a Hilbert
space, we can also form linear combinations among them, and this invariably
will bring typical quantum behaviour to the system (in a classical
commutative algebra, the only pure states are the Dirac measures on
the corresponding topological space, and linear combinations give
mixed states, which cannot be interpreted as events in that case.)
In the limit $\theta\,\longrightarrow\infty$ (we enter the macroscopic
realm there), we will take a suitable boundary condition so that we
recover the classical-like evolution curve.

$\;$

$\mathbf{Definition\;2.22}$: Consider now the real valued functions
$G(\theta)$ and $g(G(\theta))$. Then, the $\underline{\widetilde{\psi}^{\theta}}_{\overline{N}_{2}N_{3}}$
\emph{corresponding to the process} $\mathsf{e}_{\overline{N}_{1}N_{2}}\bigcup\mathsf{e}_{\overline{N}_{2}N_{3}}$
(i.e., the event $\mathsf{e}_{\overline{N}_{2}N_{3}}$, but now \emph{seen
as the endpoint of the considered path}\footnote{What this means is that, while the process is certainly $\mathsf{e}_{\overline{N}_{1}N_{2}}\bigcup\mathsf{e}_{\overline{N}_{2}N_{3}}$
and then it cannot be represented by the state $\underline{\psi}_{\overline{N}_{2}N_{3}}$,
once it happens we are only interested in its endpoint when trying
to consider the probability for the next transition after this process.
The endpoint is, of course, the elementary process $\underline{\psi}_{\overline{N}_{2}N_{3}}$,
and that's why we will say that the initial $\underline{\psi}_{\overline{N}_{1}N_{2}}$
jumps to the final $\underline{\psi}_{\overline{N}_{2}N_{3}}$, \emph{via}
$\mathsf{e}_{\overline{N}_{1}N_{2}}\bigcup\mathsf{e}_{\overline{N}_{2}N_{3}}$,
and only use $\underline{\psi}_{\overline{N}_{2}N_{3}}$ as the initial
state for the next transition.} defined as $\mathsf{e}_{\overline{N}_{1}N_{2}}\bigcup\mathsf{e}_{\overline{N}_{2}N_{3}}$,
or, equivalently, as the start of $\mathsf{e}_{\overline{N}_{3}N_{4}}$),
is given by

\[
\underline{\widetilde{\psi}^{\theta}}_{\overline{N}_{2}N_{3}}\doteq\phi_{\theta}(\underline{\psi}_{\overline{N}_{1}N_{2}})\doteq g(G(\theta))\underline{\psi}_{\overline{N}_{1}N_{2}}+G(\theta)(\underline{\psi}_{\overline{N}_{2}N_{3}}+\underline{I}),
\]
where $\phi_{\theta}$ is \emph{continuous}, with

\[
G(0)=0,\,G'(0)\neq0,\,g(0)=1.\,\blacksquare
\]

$\;$

Thus,

\[
\underline{\widetilde{\psi}^{0}}_{\overline{N}_{2}N_{3}}=\phi_{0}(\underline{\psi}_{\overline{N}_{1}N_{2}})=\underline{\psi}_{\overline{N}_{1}N_{2}},
\]
that is, if $\theta=0$, then it makes no sense to distinguish between
both events, they must be the same; furthermore, in light of this,
the process and path for the classical case, in which $\theta\,\longrightarrow\infty$,
should correspond to $\underline{\widetilde{\psi}^{\infty}}_{\overline{N}_{2}N_{3}}=\underline{\psi}_{\overline{N}_{2}N_{3}}$. 

$\;$

$\mathbf{Definition\;2.23}$: Now that we have two events, $\mathsf{e}_{\overline{N}_{1}N_{2}}$
and $\mathsf{e}_{\overline{N}_{2}N_{3}}$, we can calculate the \emph{transition
probability }between the events \emph{when following a particular
path in spacetime given by the process} $\mathsf{e}_{\overline{N}_{1}N_{2}}\bigcup\mathsf{e}_{\overline{N}_{2}N_{3}}$,
which we define, as

\[
\mathcal{P}_{\mathsf{e}_{\overline{N}_{1}N_{2}}\rightarrow\mathsf{e}_{\overline{N}_{3}N_{4}}^{start}}(\theta)^{\frac{1}{2}}\doteq(\underline{\widetilde{\psi}}_{\overline{N}_{2}N_{3}}^{\theta},\underline{\psi}_{\overline{N}_{1}N_{2}})_{\mathcal{H}_{\omega}}=g(G(\theta))\parallel\underline{\psi}_{\overline{N}_{1}N_{2}}\parallel_{\mathcal{H}_{\omega}}^{2}+
\]

\[
G(\theta)(\underline{\psi}_{\overline{N}_{2}N_{3}},\underline{\psi}_{\overline{N}_{1}N_{2}})_{\mathcal{H}_{\omega}}+G(\theta)(\underline{I},\underline{\psi}_{\overline{N}_{1}N_{2}})_{\mathcal{H}_{\omega}}
\]

\[
=g(G(\theta)).\,\blacksquare
\]

$\;$

In the classical situation, we have

\[
\mathcal{P}_{\mathsf{e}_{\overline{N}_{1}N_{2}}\rightarrow\mathsf{e}_{\overline{N}_{3}N_{4}}^{start}}^{Class.}\doteq\mid(\underline{\psi}_{\overline{N}_{1}N_{2}},\underline{\psi}_{\overline{N}_{2}N_{3}})_{\mathcal{H}_{\omega}}\mid^{2}\,\mathrm{or}\,\mid(\underline{\psi}_{\overline{N}_{1}N_{2}},\underline{\psi}_{\overline{N}_{1}N_{2}})_{\mathcal{H}_{\omega}}\mid^{2},
\]
and then we get only two possible values for the transition probability,
namely, $0$ or $1$, respectively.

$\;$

$\mathbf{Remark\;2.14}$: The map $\phi_{\theta}$ can be seen as
the equivalent, in this formalism, of (the push-forward of) a $1-$parameter
family of diffeomorphisms, and $\phi_{\theta}(\underline{\psi}_{\overline{N}_{1}N_{2}})$
can be seen as the \emph{``potential time evolution'' in the phase
space picture} for the current quantum theory. Nevertheless, the physical
interpretation is quite different. Indeed, in standard quantum physics,
a well defined (by the true Hamiltonian with respect to a background
spacetime geometry) and unique time evolution gives the state of the
system at time $t$. Here, we don't even have a ``time $t$'' to
begin with: what $\phi_{\theta}(\underline{\psi}_{\overline{N}_{1}N_{2}})$
actually represents is, for a given possible time $\theta$, a possible
time evolution itself to it, that is, in this situation, \emph{the
different time evolutions (where what varies among them is the value
of $\theta$) themselves are the ones which have different probabilities
of occurring} for the \emph{always fixed} pair of ``manifold points''/state
supports $\underline{\psi}_{\overline{N}_{1}N_{2}}$ and $\underline{\psi}_{\overline{N}_{2}N_{3}}$.
In this way, $\mathcal{P}_{\mathsf{e}_{\overline{N}_{1}N_{2}}\rightarrow\mathsf{e}_{\overline{N}_{3}N_{4}}^{start}}(\theta)^{\frac{1}{2}}$
is the probability that, when the initial event $\mathsf{e}_{\overline{N}_{1}N_{2}}$
goes to the final $\mathsf{e}_{\overline{N}_{2}N_{3}}$ (whose end
sits at time $\theta+\theta_{12}$ from $\mathsf{e}_{\overline{N}_{1}N_{2}}$:
since we have only two events here, $\theta$ can only take that value,
besides $0$, of course), it will do it via the time evolution $\phi_{\theta}(\underline{\psi}_{\overline{N}_{1}N_{2}})$.
Note that $\phi_{\theta}(\underline{\psi}_{\overline{N}_{1}N_{2}})$
\emph{doesn't} represent the state after the evolution, i.e., $\underline{\psi}_{\overline{N}_{2}N_{3}}$,
it's more than that, $\mathcal{P}_{\mathsf{e}_{\overline{N}_{1}N_{2}}\rightarrow\mathsf{e}_{\overline{N}_{3}N_{4}}^{start}}(\theta)^{\frac{1}{2}}$
is the probability that $\underline{\psi}_{\overline{N}_{1}N_{2}}$
quantum jumps/transitions to $\phi_{\theta}(\underline{\psi}_{\overline{N}_{1}N_{2}})$,
which is the state in which a given time evolution (seen as some particular
change or process) defines and in which event $\mathsf{e}_{\overline{N}_{2}N_{3}}$
has probability $1$ for ocurring, so that time actually acquires
the value $\theta$, and then, since the notion of time evolution
as something giving the state of the system at a defined time can
make sense, can be reinterpreted as an evolution more similar to the
one from standard quantum physics, in particular one which is such
that the state at time $\theta$ is $\underline{\psi}_{\overline{N}_{2}N_{3}}$$.\,\blacksquare$

$\;$

It remains to determine the functions $g,G$. For this, we will consider
the dynamics as encoded in the spectral action. The spectral action
on the \emph{NC $4-$Torus} is given by (\cite{key-9})

\[
S^{0}=\omega(F_{\mu\nu}F^{\mu\nu}),
\]
where

\[
C^{\infty}(\mathbb{T}_{\Theta}^{4})\ni F_{\mu\nu}\doteq\delta_{\mu}(A_{\nu})-\delta_{\nu}(A_{\mu})+\left[A_{\mu},A_{\nu}\right],
\]
and $C^{\infty}(\mathbb{T}_{\Theta}^{4})\ni A_{\mu}=-A_{\mu}^{*}$. 

$\;$

$\mathbf{Definition\;2.24}$: We define the following ``displaced''
version of the previous spectral action (we restrict $A_{\mu}$ to
be only from any of the two algebras of the two events in consideration,
since that's where the action of $\phi_{\theta}$ was defined) 

\[
S^{\theta}\doteq\omega(\phi_{\theta}(F_{\mu\nu}F^{\mu\nu}))=g(G(\theta))S^{0}+G(\theta).\,\blacksquare
\]

$\;$

$\mathbf{Hypothesis\;2.3}$: We impose its \emph{invariance}\footnote{Since the GNS Hilbert space of the NC $4-$Torus represents phase
space, and the elements of the algebras are vectors/states there,
then we can see $S^{0}$ acting on the events/states $\underline{\psi}_{\overline{N}_{1}N_{2}},\underline{\psi}_{\overline{N}_{2}N_{3}}$
as a functional acting on the covariant phase and expressed using
the spacetime picture (for the particular flat-metrics given by $D_{P_{N_{1}}P_{N_{2}}}$
and $D_{P_{N_{2}}P_{N_{3}}}$), in a way very analogous to the classical
Gravitational action $S_{GR}\left[g\right]=\int_{M}R(g)\boldsymbol{\mathbf{\epsilon}}$
(one would need to make a conformal perturbation of the flat metric
of the NC torus in order to have a non-zero curvature \cite{key-15},
that's why the present action $S^{0}=\omega(F_{\mu\nu}F^{\mu\nu})$
contains only a Yang-Mills term); the action of $\omega$ can be seen
as an analogous to integration. Note that we can make all these interpretations
\emph{only} thanks to the fact that we have a spacetime picture.} (because the $\phi_{\theta}$ are the analogous of physical diffeomorphisms)
on the variable $\theta$:

\[
\frac{\mathrm{d}S^{\theta}}{\mathrm{d}\theta}=0.\,\blacksquare
\]

$\;$

$\mathbf{Lemma\;2.8}$: Hypothesis 2.3 implies that $g(G(\theta))_{S^{0}}=1-\frac{1}{S^{0}}G(\theta)$.

$\;$

\emph{Proof}:

\[
\frac{\mathrm{d}S^{\theta}}{\mathrm{d}\theta}=\frac{\mathrm{d}g}{\mathrm{d}\theta}S^{0}+\frac{\mathrm{dG}}{\mathrm{d}\theta}
\]

\[
0=\frac{\mathrm{d}g}{\mathrm{d}G}\frac{\mathrm{d}G}{\mathrm{d}\theta}S^{0}+\frac{\mathrm{d}G}{\mathrm{d}\theta}\,\Longrightarrow0=\frac{\mathrm{d}g}{\mathrm{d}G}S^{0}+1\,\Longrightarrow
\]

\[
g(G(\theta))_{S^{0}}=k-\frac{1}{S^{0}}G(\theta),
\]
with $k=1$, since $G(0)=0$ and $g(0)=1$. In this way:

\[
g(G(\theta))_{S^{0}}=1-\frac{1}{S^{0}}G(\theta).\,\square
\]

$\;$

Now, consider the case for the solution $A_{\mu}^{0}=0$, then $S^{\theta}(A_{\mu}^{0})$
\emph{cannot} be invariant on $\theta$, and, in fact (recall that
$G'(0)\neq0$):

\[
S^{\theta}(A_{\mu}^{0})=G(\theta);
\]
each value of $\theta$ corresponds to a different physical situation
(roughly speaking, in the semiclassical case, the second event varies
its temporal ``distance'' with respect to the initial one as $\theta$
varies); thus, the three action values $S^{\theta_{1}}(A_{\mu}^{0}),$
$S^{\theta_{2}}(A_{\mu}^{0})$, $S^{\theta_{1}+\theta_{2}}(A_{\mu}^{0})$,
correspond to three different physical systems.

$\;$

$\mathbf{Hypothesis\;2.4}$: We propose that the \emph{dynamics} of
these systems under the $\theta$ variable varies in a \emph{homogeneous
}and\emph{ additive}\footnote{Like proper time defined as an integral over the curve is.}
fashion, that is (recall that the sum of actions corresponds to the
\emph{composition} or \emph{physical sum} of physical systems), 

\[
S^{\theta_{1}+\theta_{2}}(A_{\mu}^{0})=S^{\theta_{1}}(A_{\mu}^{0})+S^{\theta_{2}}(A_{\mu}^{0})\,\Longrightarrow
\]

\[
G(\theta_{1}+\theta_{2})=G(\theta_{1})+G(\theta_{2}).
\]
If $G$ is continuous, then this implies that

\[
G(\theta)=\alpha\theta,
\]
for some constant $\alpha=G'(0)\neq0$ (not that this is valid for
\emph{any} solution $A_{\mu}$, since $G$ only depends on $\theta$)$.\,\blacksquare$

$\;$

$\mathbf{Corollary\;2.1}$: The final form, \emph{completely determined
by the toral spectral action and our assumptions}, of the transition
probability becomes

\[
\mathcal{P}_{\mathsf{e}_{\overline{N}_{1}N_{2}}\rightarrow\mathsf{e}_{\overline{N}_{3}N_{4}}^{start}}(\theta)_{\alpha,A_{\mu}}^{\frac{1}{2}}=1-\frac{\alpha}{S^{0}(A_{\mu})}\theta.\,\square
\]

$\;$

$\mathbf{Remark\;2.15}$: From the form of the toral algebra $\mathcal{\widehat{A}}_{\overline{P}_{N_{1}}\overline{P}_{N_{2}}}$,
where $u_{1}u_{3}=e^{2\pi i\theta}u_{3}u_{1}$, it's clear that a
redefinition of $\theta$ by a displacement $\theta\,\longmapsto\theta+n$,
with $n\in\mathbb{N}$, gives an isomorphic algebra; thus, we can
restrict to $\theta\in\left[0,1\right]$. Furthermore, the map defined
by $u_{1}\,\longmapsto u_{3}$ and $u_{3}\,\longmapsto u_{1}$ establishes
an isomorphism between the toral algebra with $\theta$ and the one
with $1-\theta$; if we combine both of these results, then we get
that the family of toral algebras whose parameter $\theta$ ranges
on the subset $\left[0,\frac{1}{2}\right]$ exhausts \emph{all} the
possible \emph{non-isomorphic} toral algebras (for the $2-$dimensional
case, of course.) In this way, if we impose the \emph{only} physically
meaningful boundary condition, namely,

\[
lim_{\theta\longrightarrow\frac{1}{2}}\underline{\widetilde{\psi}^{\theta}}_{\overline{N}_{2}N_{3}}\propto(\underline{\psi}_{\overline{N}_{2}N_{3}}+\underline{I}),
\]
which is equivalent to asking\footnote{Recall that, in the classical case, $\mathcal{P}_{\mathsf{e}_{\overline{N}_{1}N_{2}}\rightarrow\mathsf{e}_{\overline{N}_{3}N_{4}}^{start}}^{Class.}\doteq\mid(\underline{\psi}_{\overline{N}_{1}N_{2}},\underline{\psi}_{\overline{N}_{2}N_{3}})_{\mathcal{H}_{\omega}}\mid^{2}=0$.}

\[
lim_{\theta\longrightarrow\frac{1}{2}}g(G(\theta))_{S^{0}(A_{\mu})}^{\alpha}=lim_{\theta\longrightarrow\frac{1}{2}}\mathcal{P}_{\mathsf{e}_{\overline{N}_{1}N_{2}}\rightarrow\mathsf{e}_{\overline{N}_{3}N_{4}}^{start}}(\theta)_{\alpha,A_{\mu}}=0,
\]
then, for \emph{each} solution $A_{\mu}$, there's a \emph{unique}
solution 
\[
g(G(\theta))_{S^{0}(A_{\mu})}^{\alpha_{\frac{1}{2}}}=1-2\theta
\]
to the \emph{displaced} dynamics, obtained by taking

\[
\alpha_{\frac{1}{2}}=2S^{0}(A_{\mu}).
\]
For this case, the graph of $\mathcal{P}_{\mathsf{e}_{\overline{N}_{1}N_{2}}\rightarrow\mathsf{e}_{\overline{N}_{3}N_{4}}^{start}}(\theta)_{\alpha_{\frac{1}{2}},A_{\mu}}$
is, of course, just the following:

\begin{figure}[H]
$\qquad\qquad\qquad\qquad\qquad\qquad\qquad$\includegraphics[scale=0.55]{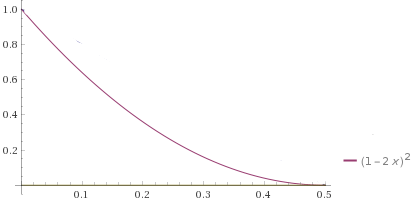}

\caption{Transition Probability for $\alpha_{\frac{1}{2}}$$.\,\blacksquare$}
\end{figure}

$\;$

$\mathbf{Remark\;2.16}$: In the \emph{unexponentiated} algebra, one
would like to have non-isomorphic algebras for \emph{any} value of
the parameter $\theta$ in the range $[0,\infty)$. Thus, in the \emph{exponentiation}
of the algebra, we made an \emph{unintended compactification}\footnote{Which, of course, is directly related to the compactness of the classical
$2-$torus $\mathbb{T}^{2}$.} of $[0,\infty)$ into $\left[0,\frac{1}{2}\right]$. To obtain the
actual domain and shape of the transition probability, we must change
the variable $\theta$ to $\vartheta\in[0,\infty)$, where the passing
from $\theta$ to $\vartheta$ is given by a \emph{bijective conformal
stretching} of $[0,\frac{1}{2})$ into $[0,\infty)$. The most sensible
option seems to be:

\[
\theta=\frac{1}{2}\mathrm{tanh}\,\vartheta,
\]
in which case

\[
\mathcal{P}_{\mathsf{e}_{\overline{N}_{1}N_{2}}\rightarrow\mathsf{e}_{\overline{N}_{3}N_{4}}^{start}}(\vartheta)_{\alpha_{\frac{1}{2}},A_{\mu}}=(1-\mathrm{tanh}\,\vartheta)^{2}
\]

\[
=e^{-4\vartheta}(1+\mathrm{tanh}\,\vartheta)^{2};
\]
now, for $\vartheta>>4$, this reduces to\footnote{Since $(1+\mathrm{tanh}\,\vartheta)^{2}\sim4,\,\forall\vartheta>>4.$}

\[
\mathcal{P}_{\mathsf{e}_{\overline{N}_{1}N_{2}}\rightarrow\mathsf{e}_{\overline{N}_{3}N_{4}}^{start}}(\vartheta)_{\alpha_{\frac{1}{2}},A_{\mu}}\sim4e^{-4\vartheta}.
\]

$\;$

Then the graph becomes:

\begin{figure}[H]
$\qquad\qquad\qquad\qquad\qquad\qquad\qquad$\includegraphics[scale=0.55]{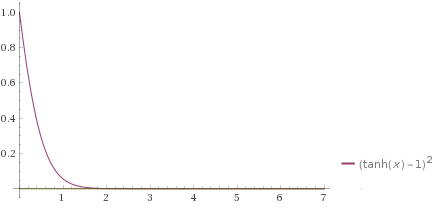}

\caption{Transition Probability for $\alpha_{\frac{1}{2}}$ (\emph{conformally
stretched})$.\,\blacksquare$}
\end{figure}

$\;$

The state of the system is given by $\Psi=\underline{\psi}_{\overline{N}_{1}N_{2}}$,
then it has \emph{non-zero} transition probability for \emph{several}
other processes; that is, upon \emph{interaction} with the surrounding
matter fields, the initial state $\underline{\psi}_{\overline{N}_{1}N_{2}}$
can make a quantum jump to the start of another event $\underline{\psi}_{\overline{N}_{n+1}N_{n+2}}$
via an elementary, ``would be compound'' process $\mathsf{e}_{\gamma_{\overline{N}_{1}\overline{N}_{n}}N_{n+1}}$,
which may even lie in a \emph{different} curve $\gamma'$ (see next
figure), and this is the physical interpretation we give to $\mathcal{P}_{\mathsf{e}_{\overline{N}_{1}N_{2}}\rightarrow\mathsf{e}_{\overline{N}_{n+1}N_{n+2}}^{start}}(\mathrm{a}(\gamma_{\overline{N}_{2}\overline{N}_{n}},P_{N_{n+1}}))_{\alpha_{\frac{1}{2}},A_{\mu}}$.
We also note that, unlike the classical case, we don't need to introduce
change in an ad-hoc manner here, since change can be seen as arising
from quantum collapse\footnote{Which, for more precision, we define as the abrupt jump from one state
to another after an interaction, if one accepts that Quantum Theory
is a complete theory, or as the abrupt change in the values, if one
doesn't accept that and, instead, introduces contextual hidden variables
(which give us the values.)} (\emph{now taken as ontologically fundamental}\footnote{This is the case, for example, in Rovelli's Relational Interpretation
\cite{key-10}, which, then, seems very well suited for this approach.} \emph{and irreducible}) after an interaction. We take the collapse
as the \emph{only} source of change and actually \emph{identify} it
with it (in other views, collapse, of course, implies change, but
the converse is not necessary; here we say they are indeed the same
thing.) Thus, the change in the classical theory actually comes from
the fundamental quantum theory, of which the former is a limit. Furthermore,
in light of this, then the argument used to show the necessity of
the collapse in QM can be now used to show the necessity of change,
which then becomes a quantum phenomena and very tied to the characteristic
non-commutativity of quantum properties.

$\;$

\begin{figure}[H]
$\qquad\qquad\qquad\qquad\qquad\qquad\qquad\qquad$\includegraphics[scale=1.5]{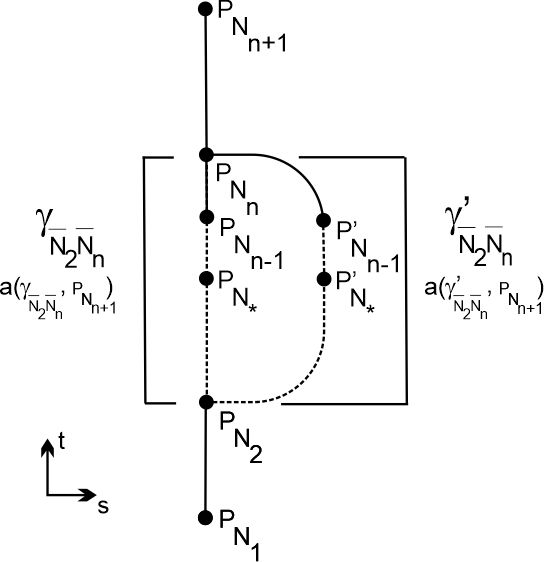}

\caption{Two Different Curves.}
\end{figure}

$\;$

In the actual physical reality, the system is constantly interacting
and its state collapsing. Therefore, its real spacetime trajectory
is something like what's illustrated in the figure below (note that
not all of the intermediate events in the curves $\gamma_{\overline{N}_{1}\overline{N}_{n}}$
are visited, i.e., not all the intermediate values of process-area
will be visited.) It's actually for this trajectory that we can define
something like 
\[
m\,\longmapsto\mathrm{a}(\gamma_{\overline{N}_{2}\overline{N}_{n(m)}},P_{N_{n(m)+1}})\,\longmapsto\mathsf{e}(\mathrm{a}(\gamma_{\overline{N}_{2}\overline{N}_{n(m)}},P_{N_{n(m)+1}}))\equiv\mathsf{e}_{\overline{N}_{n(m)+1}N_{n(m)+2}}^{start},
\]
as in the classical $t\,\longmapsto$$\tau(t)\,\longmapsto\gamma(\tau(t))$.

\begin{figure}[H]
$\qquad\qquad\qquad\qquad\qquad\qquad\qquad\qquad$\includegraphics[scale=1.5]{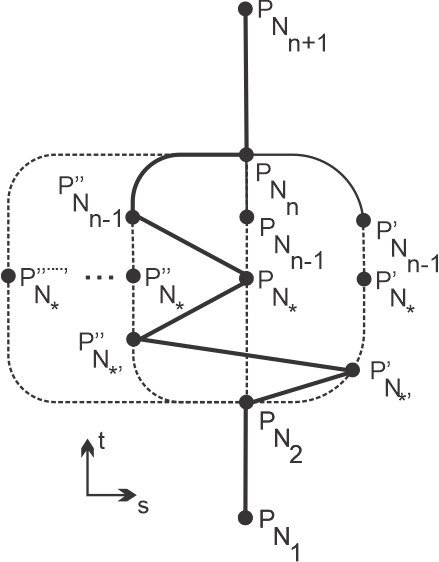}

\caption{Real Spacetime Trajectory of the System.}
\end{figure}

$\;$

$\mathbf{Remark\;2.17}$: The so-called \emph{Problem of Time} can
be resolved here simply by calculating the values of the properties
of interest (say, the spatial curvature) on each of the nodes of $\mathsf{e}(\mathrm{a}(\gamma_{\overline{N}_{2}\overline{N}_{n(m)}},P_{N_{n(m)+1}}))$.
Of course, since quantum collapse is random, we cannot predict with
certainty what's the trajectory that the system will take, and thus
the precise time evolution of the value for the property being considered.
The only thing we can do is to calculate the probability for each
possible time evolution of the value, by setting:

\[
\mathcal{P}(m)\doteq\prod_{k=1}^{m}\mathcal{P}_{n(k),n'(k)}^{spatial}\mathcal{P}_{\mathsf{e}_{\overline{N}_{n(k)}N_{n(k)+1}}\rightarrow\mathsf{e}_{\overline{N}_{n'(k)+1}N_{n'(k)+2}}^{start}}(\mathrm{a}(\gamma_{\overline{N}_{n(k)+1}\overline{N}_{n'(k)}},P_{N_{n'(k)+1}}))_{\alpha_{\frac{1}{2}},A_{\mu}},
\]
where only the events in $\mathsf{e}(\mathrm{a}(\gamma_{\overline{N}_{2}\overline{N}_{n(m)}},P_{N_{n(m)+1}}))$
are considered for the product$.\,\blacksquare$

$\;$

$\mathbf{Remark\;2.18}$: Now, since $\mathcal{P}_{\mathsf{e}_{\overline{N}_{1}N_{2}}\rightarrow\mathsf{e}_{\overline{N}_{n+1}N_{n+2}}^{start}}(\mathrm{a}(\gamma_{\overline{N}_{2}\overline{N}_{n}},P_{N_{n+1}}))_{\alpha_{\frac{1}{2}},A_{\mu}}$
falls exponentially, then the events $\mathsf{e}_{\gamma_{\overline{N}_{2}\overline{N}_{n}}N_{n+1}}$
for \emph{large} $n$ are very unlikely to happen. This means that
$\mathsf{e}_{\overline{N}_{1}N_{2}}$ tends to go to a $\mathsf{e}_{\overline{N}_{n+1}N_{n+2}}^{start}$
which is closer to it in proper time separation. Physically, this
means that time advances as a succession of instants which are very
close to each other in proper time distance and in which the duration
of the instants themselves is very small. Thus, at the macroscopic
scale, this is perceived as a succession of instants, each of duration
zero, which forms a continuum whose subsets have finite duration,
and which monotonically increases (since, upon change, almost all
intermediate steps are visited, and, thus, the system can interact
with whatever thing that resides at those steps), that is, the classical
picture of time. The quantum system at an event is surrounded by a
dispersion cloud of events which can visit next, and the classical
proper time is just some average $<\tau>$ of that dispersion, and
the actual quantum transitions measure how much the actual quantum
time \emph{deviates} from this average $<\tau>$$.\,\blacksquare$

$\;$

$\mathbf{Remark\;2.19}$: Also in relation to this, this point of
view may also shed some light on the so-called \emph{Measurement Problem}
of standard QM, understood as the \emph{impossibility} to \emph{explain}
the collapse of the wavefunction in terms of the standard Schrödinger
time evolution (assuming that the collapse gives definite values and
that standard QM is complete.) Indeed, in our view, the standard classical
time, which, among other things, is used as the external time parameter
to define the Schrödinger time evolution, is only an emergent feature
at the macro level, and fuelled at the micro, fundamental level by
an irreducible collapse. Thus, the Schrödinger time evolution will
\emph{break} when something from the fundamental level \emph{leaks}
to the macroscopic level. And, of course, this is precisely the situation
in a quantum measurement, when the state collapses when certain two
systems interact: the classical time will never be able to explain
this process since the latter intervenes precisely in making possible
the quantum, and therefore also the classical, time. Finally, one
can also may be able to avoid spacetime singularities (the ``\emph{Singularity
Problem}'' of classical GR) in this approach thanks to the discretization
of time and the mentioned ``tunneling'' effect. Indeed, the discretization
eliminates the possibility for a property to acquire values which
are finite yet arbitrary large as one approaches an event separated
by a finite amount of proper time (either to the past or future) with
respect to the initial one, since there are only a finite number of
other events between them, thus the property reaches a finite maximum
value, possibly at the final event; furthermore, if the property is
not defined for an event, then, of course, it cannot form part of
the spacetime, yet, due to the possibility of tunneling, even if the
classical spacetime is inextendible, there still exists a non-zero
transition probability for the system to tunnel to an extended quantum
spacetime, that is, beyond the ``boundary'' of the classical spacetime
defined by the singularity$.\,\blacksquare$

$\;$

\subsection*{{\normalsize{}3. Discussion}}

$\;$

In CLQG (Covariant LQG \cite{key-13}), given the Hilbert spaces and
quantum algebras at the boundary, it seems one can only calculate
things like (quantum) properties of the induced spatial $3-$d metric
(areas, volumes, curvature) and the extrinsic curvature of the hypersurface,
and not quantum durations (for example, in the analysis done for the
extrinsic coherent states, the area and the extrinsic curvature peak
at those states, but time is introduced as an \emph{already} peaked
external parameter.) There are some extensions that introduce timelike
boundaries, but, for a time evolution, we need to be able to introduce
durations also in the \emph{interior} of the process; furthermore,
the commutation relations for variables whose weight functions have
a space\emph{time} support, and in regions in spacetime that are such
that one is in the causal future of the other, are non-trivial (this
is the important result of the so-called covariant Poisson brackets),
and are not accounted for by the \emph{part} of the commutation relations
that arise \emph{purely} due to the structure (kinematical in nature)
of the group in the Yang-Mills-like/connection variables used in these
formulations \cite{key-14}, since the dependence over the causality
comes from the hyperbolic character of the dynamics and the mentioned
issue with the supports, and then it's indifferent if we take the
Yang-Mills/connection group to be $SU(2)$ (spacelike boundary) or
$SU(1,1)$ (timelike boundary), that is, this causal part will be
present even if the field variable were a scalar field; as argued
in the paper, and from relational considerations, we consider this
a key issue in the quantization of duration, which makes it quite
different from the quantization of the spatial metric and that it
cannot thus be completely obtained by tweaks on the already existing
methods for the latter. In particular, the extensions of CLQG give
the area of a timelike triangle for vectors (one timelike and the
other spacelike) whose origin is at the same vertex (evidently, the
causal part is irrelevant, since the origin of both vectors is the
same point, and then only the group part enters to play there), while
TQG describes the same area but from the point of view of the causal
connection between the initial and enpoint of the timelike vector
(thus, the causal part is relevant for that); both refer to a same
event and as a finite process, but TQG gives a more dynamical characterization
of it while CLQG gives\footnote{Note that the dynamical transition in CLQG between the initial and
final points of the previous timelike vector is not the same thing
as the process described by TQG between them: the transition in CLQG
is actually a transition between two processes, the latter ones each
of kinematical area given by the triangles at the points in question,
while the events in TQG describe dynamically (via the causal relation
between the initial and final points) the processes corresponding
to \emph{each} triangle. The dynamical transitions in CLQG cannot
describe this latter dynamical characterization for each individual
triangle because they are transitions from state to state, triangle
to triangle. The transition described in CLQG corresponds to the transition
between events in TQG, but where each theory describes different aspects
of it (in CLQG, is the dynamics between the kinematics of each event,
while in TQG is the dynamics between the dynamics of each.) We would
say that it's actually TQG the theory that gives actual physical entity
to those events since it describes them as dynamics.} its internal kinematical degrees of freedom. Thus, the corresponding
quantum algebra for duration is a composition of the group and causal
aspect of the classical Poisson bracket. From this causal part, unique
to TQG, one gets an \emph{universal} decay for the transition that
explains the features of duration at the macro scale (this decay is
always there and is independent of the details of the other parts
of the transitions); this part of the transition for duration is ignored
(or considered to have already peaked) in treatments that rely heavily
in coherent states adapted only to the group part.

$\;$

In classical GR, the \emph{full} form of a solution is $g_{ab}=n_{a}n_{b}+h_{ab}$,
with $n^{a}=\frac{1}{N}(t^{a}+N^{a})$, where $N$ and $N^{a}$ are,
respectively, the lapse and shift (which also satisfy $N=-g_{ab}n^{a}t^{a}$
and $N^{a}=h_{b}^{a}t^{b};$ in this way, $N$ is interpreted as the
``rate of proper time with respect to coordinate time $t$ as one
moves normaly to the hypersurfaces of the foliation''), so that the
proper time becomes $\tau=\int\sqrt{N^{2}-h_{ab}N^{a}N^{b}}\,\mathrm{d}t$.
Now, it's also known that the induced metric and the extrinsic curvature
(the data given by the boundary Hilbert spaces and algebras of CLQG)
determine a point in phase space if we were in the classical theory.
But this is not enough to generate a solution, one also needs to specify
the lapse and shift $N$ and $N^{a}$ (which measure how the time
function/coordinate $t$ of the foliation $\varSigma_{t}$ of the
manifold interacts with the metric information.) Furthermore, from
the previous formulas we can see that the lapse and shift $N$ and
$N^{a}$ are precisely the things that encode the time part/time evolution
of the spacetime\emph{ }behaviour of the solution, as well as the
proper time (we just can't have one without the other.) Therefore,
in order to describe time evolution properly, we must introduce some
extra information to CLQG, which is given by an algebra that can econde
the non-trivial commutation relations between causally connected regions
in spacetime in the interior of the process (and find a way to describe
metric properties related to it.) But, even if we have this, the transition
amplitude $W(x_{i},x_{f})$ is not useful for obtaining the amplitude
for the transitions between quantum states of duration, since it's
based on the path integral, which, as one can see in the canonical
framework, involves integration over all lapse functions $N$, and
then information about the duration part of a \emph{particular} solution
cannot be taken as being prescribed from there. Thus, the dynamical
transitions for duration must also be provided by different means.
TQG provides both of these necessary ingredients for the adequate
description of time evolution in QG (the areas of the elementary processes
in the paper can be interpreted as the information provided by $N$
and $N^{a}$ in the classical case.)

$\;$

The probabilty amplitude $\mid W(x_{i},x_{f})\mid^{2}$ of CLQG should,
perhaps, be interpreted as what is denoted as $\mathcal{P}_{n(k),n'(k)}^{spatial}$
here. On the other hand, TQG, \emph{by providing an actual description
of time duration in QG}, is what actually justifies the physical interpretation
of $W(x_{i},x_{f})$ as a transition amplitude in CLQG, the latter
is a thing which has often been criticized (\cite{key-2}) due to
the integration over all lapse functions on the path integral: if
we delegate the time part to TQG, then this integration can be seen
only as a device of the calculation, since no contradiction arises
because the path integrals are being applied to states that only describe
space and to obtain the pure space part of the transition amplitude.

$\;$

Thus, we actually see CLQG and TQG as \emph{complementary} theories.

$\;$

\subsection*{{\normalsize{}4. Conclusions}}

$\;$

We made a schematic partial quantization of the temporal part in line
with the mentioned ideas. For doing that, we had to introduce a new
formalism for QG, which we call TQG, since it's heavily based on the
NC Tori. This allowed us to obtain numerous insights about the nature
of time, like its discretization, its regular pace at the macroscopic
scale, a solution to the Problem of Time, and a connection with the
Measurement Problem of QM.

$\;$

TQG is not that much a theory of quantum gravity, but a necessary
part of it. In particular, it provides a basic framework on which
to discuss within such a theory any aspect related to time, time evolution,
and duration. In this proposal, the part of quantum phase space that
deals with these aspects is modeled by an $n-$dimensional noncommutative
Torus, where each state on a NC-2-subTorus there corresponds to an
elementary and irreducible quantum process in a spacetime picture;
the NC-Torus arises here because for, say, three (of four) elements
that are causally related, one needs three generators that do not
commute among each other (this comes from the canonical quantization
of the Poisson brackets of phase space properties with weight functions
having causally connected supports), and this cannot be accomodated
just in the product of two NC-2-Torus (that model an individual quantum
elementary process), but one needs a missing third non-commutation
between these two, so to speak, and this can be done in the general
$3$ and $4-$NC-Torus. This type of phase space hasn't been discussed
in any of the current proposal for quantum gravity theories, despite
the fact that often some talk is informally given (\cite{key-15,key-4})
about the behaviour of quantum time (like its hypothetical quantum
nature, with superpositions, etc.), that, really, can only be properly
justified if one has a model for this phase space.

$\;$

\end{document}